\documentclass[12pt]{article}

\usepackage{amsmath,amssymb}
\usepackage{slashed}
\usepackage{xcolor} 
\usepackage{graphicx}
\usepackage{url}
\usepackage{cite}
\usepackage{pdflscape}

\newcommand{\fref}[1]{Fig.~\ref{fig:#1}} 
\newcommand{\eref}[1]{Eq.~\eqref{eq:#1}}

\newcommand{\aref}[1]{Appendix~\ref{app:#1}}
\newcommand{\sref}[1]{Section~\ref{sec:#1}}
\newcommand{\cref}[1]{Chapter~\ref{ch:.#1}}
\newcommand{\tref}[1]{Table~\ref{tab:#1}}

\newcommand{\nnl}{\nonumber \\}

\newcommand{\beq}{\begin{equation}} 
\newcommand{\eeq}{\end{equation}} 
\newcommand{\ba}{\begin{array}}  
\newcommand{\ea}{\end{array}} 
\newcommand{\bea}{\begin{eqnarray}}  
\newcommand{\eea}{\end{eqnarray} }  
\newcommand{\be}{\begin{eqnarray}}  
\newcommand{\ee}{\end{eqnarray} }  
\newcommand{\bal}{\begin{align}}
\newcommand{\eal}{\end{align}}   
\newcommand{\bi}{\begin{itemize}}  
\newcommand{\ei}{\end{itemize}}  
\newcommand{\ben}{\begin{enumerate}}  
\newcommand{\een}{\end{enumerate}}  
\newcommand{\bc}{\begin{center}}
\newcommand{\ec}{\end{center}} 
\newcommand{\bt}{\begin{table}}
\newcommand{\et}{\end{table}}  
\newcommand{\btb}{\begin{tabular}}
\newcommand{\etb}{\end{tabular}}  
\newcommand{\bvec}{\left ( \ba{c}}
\newcommand{\evec}{\ea \right )}

\newcommand{\moller}{M\o ller }

\newcommand{\cO}{{\mathcal O}} 
 
\newcommand{\cL}{{\mathcal L}} 
 
\newcommand{\cM}{{\mathcal M}}

\newcommand{\gev}{\mathrm{GeV}}

\def\hc{{\rm h.c.}} 

\begin{document}

\vspace*{-2cm}
\begin{flushright}
\vspace*{2mm}
\today
\end{flushright}

\begin{center}
\vspace*{15mm}

\vspace{1cm}
{\large \bf 
Model independent constraints on four-lepton operators
} \\
\vspace{1.4cm}

{Adam Falkowski$\,^{a}$,  and  Kin Mimouni$\,^{b}$.}

 \vspace*{.5cm}
${}^a$ Laboratoire de Physique Th\'{e}orique, Bat.~210, Universit\'{e} Paris-Sud, 91405 Orsay, France.
\\
${}^b$ Institut de Th\'{e}orie des Ph\'{e}nom\`{e}nes Physiques, EPFL, Lausanne, Switzerland. 

\vspace*{.2cm} 

\end{center}

\vspace*{10mm} 
\begin{abstract}\noindent\normalsize

We obtain constraints on 4-lepton interactions in the effective field theory with dimension-6 operators. 
To this end, we combine the experimental input from  Z boson measurements in LEP-1, W boson mass and decays, muon and tau decays,  lepton pair production in LEP-2,  neutrino scattering on electrons, and parity violating electron scattering.   
The analysis does not rely on  any assumptions about the flavor structure of the dimension-6 operators. 
Our main results are the confidence intervals for Wilson coefficients of 16 lepton-flavor conserving four-lepton operators, together with the full correlation matrix.  
Consequences for leptophilic models beyond the Standard Model are discussed.

\end{abstract}

\vspace*{3mm}

\newpage
 
\section{Introduction}
\label{sec:intro}

An effective field theory~(EFT) provides a model-independent framework to  characterize new physics beyond the Standard Model~(SM).
It gives an adequate description of physical processes in current and past experiments if the new particles are much heavier than the weak scale. 
 Then new physics effects can be represented,  without introducing new degrees of freedom,  by operators with canonical dimensions $D>4$ added to the SM Lagrangian. 
Assuming lepton number conservation, leading effects are expected to originate from $D$=6 operators~\cite{Buchmuller:1985jz}.
 It is important to understand and describe, with minimal theoretical assumptions, the existing experimental constraints on these operators. 
While many $D$=6 operators can be probed at the LHC collider, others can be better constrained by previous experiments at high and low energies. 

The program of systematic characterization of experimental constraints on the Wilson coefficients $D$=6 operators was pioneered in Ref.~\cite{Han:2004az},  and later continued e.g. in Refs~\cite{Han:2005pr,Barbieri:2004qk,Grojean:2006nn,Cacciapaglia:2006pk,Pomarol:2013zra,Elias-Miro:2013mua,Dumont:2013wma,Chen:2013kfa,deBlas:2013qqa,Willenbrock:2014bja,Gupta:2014rxa,Masso:2014xra,deBlas:2014ula,Ciuchini:2014dea,Ellis:2014huj,Falkowski:2014tna,delAguila:2014soa,Corbett:2015ksa,Efrati:2015eaa,Gonzalez-Alonso:2015bha,Buckley:2015nca,deBlas:2015aea,Falkowski:2015fla,delAguila:2015vza,Wells:2015eba,Berthier:2015gja,Ellis:2015sca,Englert:2015hrx}.  
Most often, due to a huge number of independent $D$=6 operators \cite{Grzadkowski:2010es,Alonso:2013hga}, additional assumptions about  the flavor structure of fermionic operators are made.    
The exception was Ref.~\cite{Efrati:2015eaa} where a completely generally flavor structure was allowed.  
That work derived electroweak precision constraints on the subset of $D$=6 operators that yields the so-called {\em vertex corrections} to $Z$ and $W$ boson interactions with the SM fermions.  
Only observables that are not affected at leading order by four-fermion operators were considered, so as to separate the analysis of vertex corrections from that of (more numerous) four-fermion operators.
Assuming that $D$=6 operators give the dominant new physics contributions to observables, 
and that loop suppressed new physics contributions are sub-dominant,  
20 flavor preserving vertex corrections can be simultaneously constrained. 
A Gaussian likelihood function in the space of  the vertex corrections was given,  which can  be used to constrain any particular model beyond the SM  predicting a more restricted pattern of vertex corrections in the low-energy EFT.

In this paper we extend the analysis of Ref.~\cite{Efrati:2015eaa} so as to also obtain constraints on some $D$=6 four-fermion operators. 
This step is important in order to probe a larger class of theories than what can be achieved using vertex corrections only.
For example, if the UV theory contains  vector boson coupled to the SM fermions but not mixing with the Z or W boson, then it gives rise  {\em only} to 4-fermion operators in the low-energy EFT.  
Again, the main goal is to derive experimental constraints without assuming anything about the flavor structure of $D$=6 operators.  
We focus on {\em four-lepton} operators, leaving four-fermion operators with quarks for future publication. 
These operators were probed most precisely by lepton pair production in LEP-2 and W mass measurements. 
However, these observables leave several unconstrained directions in the space of four-lepton operators. 
Therefore, we also include in our analysis low-energy precision experiments, such as neutrino scattering on electrons, parity violating electron scattering, and leptonic decay of muons and taus. 
This way we are able  to simultaneously constrain 16 linear combinations of 27 Wilson coefficients of lepton-flavor conserving four-lepton operators, in addition to the previous  constraints on the vertex corrections. 
The correlation matrix  is given,  which allows one to reconstruct the full likelihood function. 
Given that,  our results can be used to constrain large classes of   models that give rise to an arbitrary  pattern of  vertex corrections and 4-lepton operators in the low-energy EFT.  

The paper is organized as follows. 
In \sref{formalism} we lay out our formalism where the effects of $D$=6 operators are parametrized by vertex corrections and four-fermion operators. 
In \sref{exp} we discuss  how relevant  experimental observables are affected by these  EFT parameters. 
In \sref{fit} we give the confidence intervals for the  leptonic vertex corrections and a subset of  lepton-flavor conserving four-lepton operators, together with the full correlation matrix (relegated to \aref{correlation}).   
In \sref{models} we discuss the consequences of our general analysis for particular models with new leptophilic particles.
In \aref{uni} we discuss the relationship between our formalism and the more familiar  oblique parameters,  and in \aref{lfv} we review current constraints on lepton-flavor violating operators.

\section{Formalism}
\label{sec:formalism}

We first summarize our conventions. 
The $SU(2) \times U(1)$ gauge couplings of the SM are denoted by $g_L$, $g_Y$. The photon coupling strength is  $e = g_L s_\theta$, where $s_\theta = g_Y/\sqrt{g_L^2 + g_Y^2}$ is the weak mixing angle. The vacuum expectation value (VEV) of the SM Higgs field $H$ is 
$\langle H^\dagger H \rangle = v^2/2$.  
The SM fermions are written using  the two-component spinor notation, and we follow the conventions of  Ref.~\cite{Dreiner:2008tw}.
The left- and right-handed charged leptons are denoted by $e_I = (e,\mu,\tau)$, $e_I^c = (e^c,\mu^c,\tau^c)$, while the neutrinos are denoted by  $\nu_I=  (\nu_e,\nu_\mu,\nu_\tau)$, where  $I = 1\dots 3$ is the flavor index. 
We work in the basis where these fermions fields are mass eigenstates. 

We consider an EFT  with the Lagrangian    
\beq
\label{eq:leff}
\cL_{\rm eff}  = \cL_{\rm SM}  +   {1\over v^2} \cL^{D=6}, \qquad   {\cal L}^ {D=6} = \sum_i c_i O_{D= 6,i}. 
\eeq 
Here $\cL_{\rm SM}$ is the SM Lagrangian, and $O_{D=6,i}$ is a complete basis of $SU(3) \times SU(2) \times U(1)$ invariant  operators of canonical dimension $D$=6 constructed out of the SM fields. 
The Wilson coefficients $c_i$ are formally $\cO(v^2/\Lambda^2)$ where $\Lambda$ is the scale of new physics that sets the EFT expansion. 
We assume here that  $O_{D=6,i}$ conserve the baryon and lepton number. 
Below we also assume that individual $U(1)_e \times U(1)_\mu \times U(1)_\tau$ lepton numbers are conserved; see \aref{lfv} for the discussion of lepton flavor violating operators.
In \eref{leff} we omit higher-dimensional operators with $D>6$.
While these are always present in an EFT derived as  a low-energy description of specific UV models,  throughout this paper we will assume that their contribution to the relevant observables is negligible.
This assumption is generically true when the scale $\Lambda$ of new physics is much larger than the electroweak scale $v$, since the Wilson coefficients of higher dimensional operators are formally  
$\cO(v^4/\Lambda^4)$ or smaller.    
Consequently, our analysis will be performed at $\cO(\Lambda^{-2})$, that is to say, we will take into account the corrections to observables that are linear in $c_i$, and ignore quadratic corrections which are formally  $\cO(\Lambda^{-4})$. 
We do not impose any constraints on the Wilson coefficients $c_i$. 
In particular, all $D$=6 operators can be simultaneously present in the Lagrangian, 
and the Wilson coefficients of fermionic operators  can be flavor dependent.       

Rather than parametrizing the theory space by the Wilson coefficients of $D$=6 operators, 
we find it more convenient to work directly with parameters describing interactions of mass eigenstates  after electroweak symmetry breaking.  
Our formalism follows that in Refs.~\cite{Gupta:2014rxa,HB}. 
Without any loss of generality, $\cL_{\rm eff}$ can be brought to a form where the  kinetic terms of all mass eigenstates are  diagonal and canonically normalized. 
Then the quadratic Lagrangian for the electroweak gauge boson and lepton mass eigenstates is given by 
\bea 
\label{eq:Leff_lkin}
\cL_{\rm eff}^{\rm kin} &= & -{1 \over 2} W_{\mu \nu}^+  W_{\mu \nu}^- -   {1 \over 4} Z_{\mu \nu} Z_{\mu \nu} 
-   {1 \over 4} A_{\mu \nu} A_{\mu \nu}  
+ {g_L^2 v^2 \over 4} \left (1+ \delta m \right )^2  W_{\mu}^+  W_{\mu}^-  
+ {(g_L^2 + g_Y^2) v^2  \over 8 } Z_\mu Z_\mu 
\nnl  & + & 
 i \bar e_I \bar \sigma_\mu \partial_\mu e_I +  i \bar \nu_I \bar \sigma_\mu \partial_\mu \nu_I
 + i    e^c_I \sigma_\mu \partial_\mu \bar e^c_I . 
\eea  
Here, $\delta m$ parametrizes the relative correction to the W boson mass that may arise in the presence of $D$=6 operators. 
By construction, there is no correction to the $Z$ boson mass: a possible shift due to  $D$=6 operators has been absorbed into the definition of the electroweak parameters $g_L$, $g_Y$ and $v$.     
For the sake of our analysis we need to define the interactions of leptons with the SM gauge fields in the effective Lagrangian: 
\bea
\label{eq:lvll}
\cL_{\rm eff}^{v \ell \ell} & =  & 
 - e A_\mu(\bar e_I \bar \sigma_\mu e_I  + e^c_I \sigma_\mu  \bar e^c_I)   
 + {g_L \over \sqrt 2}\left [ W_\mu^+  \bar \nu_I \bar \sigma_\mu (1 +  \delta g^{W e_I}_L  ) e_I+  \hc \right ]
 + \sqrt{g_L^2 + g_Y^2} Z_\mu 
\nnl &\times&       \left [
\bar \nu_I \bar \sigma_\mu \left (  {1 \over 2 }  + \delta g^{Ze_I}_L +  \delta g^{W\ell_I}_L \right ) \nu_I
+ \bar e_I \bar \sigma_\mu \left ( - {1 \over 2 }  +  s^2_\theta  + \delta g^{Ze_I}_L\right ) e_I 
+  e^c_I \sigma_\mu \left (s^2_\theta  + \delta g^{Ze_I}_R \right )\bar e^c_I  \right ],
\nnl 
\eea
Here, the effects of $D=6$ operators are parameterized by the vertex corrections $\delta g$. 
All $\delta g$'s in \eref{lvll} are independent parameters,  which in general may depend on the lepton flavor.
By construction, there is no vertex corrections to photon interactions.   
The parameters $\delta g$ can be related by a linear transformation to Wilson coefficients of $D$=6 operators in any particular basis, see Ref.~\cite{HB} for a map to popular bases used in the literature.
Therefore, $\delta g$'s are $\cO(\Lambda^{-2})$ in the EFT expansion.    
Note that the vertex corrections to neutrino interactions with $Z$  in \eref{lvll}  are expressed by the other vertex corrections:  $ \delta g^{Z\nu_I}_L = \delta g^{Ze_I}_L +  \delta g^{W e_I}_L$.  
This relation is a consequence of the linearly realized SM gauge symmetry and the absence of operators with $D>6$ in the Lagrangian, and holds independently of the basis of $D$=6 operators employed in \eref{leff}.

\begin{table}
\bc
\begin{tabular}{c|c}
\hline 
One flavor ($I=1\dots 3$) & Two flavors ($I < J =1\dots 3$) \\ 
\hline 
$ [O_{\ell \ell}]_{IIII} = {1\over 2} (\bar \ell_I\bar \sigma_\mu \ell_I)  (\bar \ell_I \bar \sigma_\mu \ell_I)$
 & $ [O_{\ell \ell}]_{IIJJ}  =  (\bar \ell_I\bar \sigma_\mu \ell_I)  (\bar \ell_J \bar \sigma_\mu \ell_J) $ 
 \\ 
  & $[O_{\ell \ell}]_{IJJI} = (\bar \ell_I \bar \sigma_\mu \ell_J)  (\bar \ell_J \bar \sigma_\mu \ell_I)  $
 \\   
$ [O_{\ell e}]_{IIII} =  (\bar \ell_I\bar \sigma_\mu \ell_I)  (e_I^c  \sigma_\mu \bar e_I^c) $ &
$ [O_{\ell e}]_{IIJJ}  =  (\bar \ell_I\bar \sigma_\mu \ell_I)  (e_J^c  \sigma_\mu \bar e_J^c)$
 \\
 &  $[O_{\ell e}]_{JJII}  =  (\bar \ell_J \bar \sigma_\mu \ell_J)  (e_I^c  \sigma_\mu \bar e_I^c)$
  \\
 &  $[O_{\ell e}]_{IJJI}  =  (\bar \ell_I \bar \sigma_\mu \ell_J)  (e_J^c  \sigma_\mu \bar e_I^c)$
\\ 
 $ [O_{e e}]_{IIII} =   {1\over 2} (e_I^c  \sigma_\mu \bar e_I^c)   (e_I^c  \sigma_\mu \bar e_I^c) $ &
 $ [O_{e e}]_{IIJJ}  =   (e_I^c  \sigma_\mu \bar e_I^c)   (e_J^c  \sigma_\mu \bar e_J^c) $
 \end{tabular}
\ec 
\label{tab:4l}
\caption{The full set of lepton flavor conserving 4-lepton operators in the $D$=6 EFT Lagrangian.}
\end{table}

The main focus of this paper is on the lepton-flavor conserving 4-lepton operators in \eref{leff}   summarized in \tref{4l}. 
Overall, there is $3 \times 3 + 3 \times 6 = 27$ such operators.  
Three of those,  denoted $[O_{\ell e}]_{IJJI}$,  are complex, in which case the corresponding Wilson coefficient is complex, and the Hermitian conjugate operator is included in \eref{leff}.
The goal of this paper  is to derive simultaneous constraints on the Wilson coefficients of (as many as possible) 4-lepton operators and the leptonic vertex corrections in \eref{lvll}.    
In our framework, the remaining parameter introduced above - the W mass correction $\delta m$ in \eref{Leff_lkin} -  is related to the leptonic vertex corrections and one 4-lepton operators \cite{HB}:  
\beq
\label{eq:HB_dm}
\delta m = {\delta g^{We}_L + \delta g^{W\mu}_L  \over 2}  -  {[c_{\ell \ell}]_{1221} \over 4}. 
\eeq 
Again, this relation is a consequence of the linearly realized SM gauge symmetry and the absence of operators with dimensions greater than 6.  
It also ensures that the Fermi constant $G_F$ measured in muon decays is given at tree-level 
by $G_F = 1/\sqrt {2} v^2$.  
This way, the tree-level relations between the electroweak parameters  $g_L$, $g_Y$ and $v$ and the input observables $\alpha_{\rm em}$, $m_Z$ and $G_F$ are the same as in the SM.

\section{Experimental Input}
\label{sec:exp}

In this section we discuss observables that will allow us to place constraints on 4-lepton operators and leptonic vertex corrections. 

\subsection{Z- and W-pole observables}

The leptonic vertex corrections in \eref{lvll} can be probed by measurements of leptonic decays of on-shell Z and W bosons. 
Precise measurements of observables ultimately related to various Z and W partial decay widths  were performed in LEP-1 (Z) \cite{ALEPH:2005ab}  and LEP-2 (W)  \cite{Schael:2013ita}.
The dependence of these observables on leptonic and quark vertex corrections is correlated. 
On the other hand, the dependence on four-fermion operators is suppressed by $\Gamma_V/m_V$ and can be neglected \cite{Han:2004az}. 
Simultaneous constraints on all flavor-preserving vertex corrections were derived recently in   Ref.~\cite{Efrati:2015eaa}, and we use directly these results.   
Marginalizing the likelihood over the quark vertex corrections, 
the constraints on the leptonic ones are given by: 
\beq
\label{eq:dgl}
\bvec 
\delta g^{We}_L \\ 
\delta g^{W\mu}_L \\ 
\delta g^{W\tau}_L \\ 
\delta g^{Ze}_L \\ 
\delta g^{Z\mu}_L \\ 
\delta g^{Z\tau}_L \\ 
\delta g^{Ze}_R  \\ 
\delta g^{Z\mu}_R \\ 
\delta g^{Z\tau}_R
\evec 
= 
\bvec
 -1.00 \pm 0.64 \\   -1.36  \pm 0.59 \\  1.95 \pm 0.79 \\
 -0.026 \pm 0.028 \\   0.01  \pm 0.11 \\  0.016  \pm 0.058 \\  
 -0.037 \pm 0.027 \\   0.00  \pm 0.13 \\  0.039  \pm 0.062
\evec 
 \times 10^{-2}, 
\eeq 
with the correlation matrix 
\beq
\rho = 
\left(
\begin{array}{ccccccccc}
 1. & -0.12 & -0.63 & -0.1 & -0.03 & 0.01 & 0.07 & -0.06 & -0.04 \\
 . & 1. & -0.56 & -0.11 & -0.04 & 0.01 & 0.08 & -0.06 & -0.04 \\
 . & . & 1. & -0.1 & -0.03 & 0.01 & 0.07 & -0.05 & -0.04 \\
 . & . & . & 1. & -0.1 & -0.07 & 0.17 & -0.05 & 0.03 \\
 . & . & . & . & 1. & 0.07 & -0.06 & 0.9 & -0.04 \\
 . & . & . & . & . & 1. & 0.02 & -0.03 & 0.41 \\
 . & . & . & . & . & . & 1. & -0.08 & -0.04 \\
 . & . & . & . & . & . & . & 1. & 0.04 \\
 . & . & . & . & . & . & . & . & 1. \\
\end{array}
\right).
\eeq 
Note that {\em all} leptonic vertex corrections are strongly constrained by the data in a model-independent way. 
In particular, the constraints on charged leptons couplings to Z (dominated by LEP-1) are at a  per-mille level,  while the constraints on lepton couplings to W  (dominated by LEP-2) are at a  percent level. 

\subsection{W mass}

The W boson mass was  measured very precisely at LEP-2 and the Tevatron. 
We use  the result from Ref.~\cite{Group:2012gb},   $m_{W}  =  (80.385 \pm 0.015)$~GeV, 
where the SM prediction is  $m_{W}  = 80.364$~GeV.    
This trivially translates into the constraint on the parameter $\delta m$ in \eref{Leff_lkin},  
\beq
\label{eq:dm}
\delta m = \left ( 2.6  \pm  1.9 \right ) \times 10^{-4}\, .
\eeq 
By virtue of \eref{HB_dm}, this result constrains a combination of leptonic vertex correction and one four-lepton operator.

\subsection{Fermion pair production in LEP-2}
\label{sec:lep2}

The LEP-2 experiment measured differential cross sections for the  processes $e^+ e^- \rightarrow \ell^+ \ell^-$, $\ell = e,\mu,\tau$ at energies above the Z boson resonance.  
Away from the Z-pole, these processes probe not only Z couplings to leptons but also 4-lepton operators, 
and the effect of the latter increases with increasing center-of-mass energy.    

Let us first focus on the processes $e^- e^+ \rightarrow \mu^- \mu^+$ 
($e^- e^+ \rightarrow \tau^- \tau^+$ is analogous). 
For the experimental input, we will use the total cross-sections and forward-backward asymmetries  measured at 12 different center-of-mass energies between $\sqrt s \approx$ 130~GeV and $\sqrt s\approx$209~GeV~\cite{Schael:2013ita}. 
We are interested in $\cO(\Lambda^{-2})$ corrections to these observables from $D$=6 operators, 
which translates to linear corrections in the vertex corrections and Wilson coefficients of 4-fermion operators (i.e. the interference term between SM and new physics). 
At that order, the observables  are affected by 5 four-leptons operators $[O_{\ell\ell}]_{1122}$, $[O_{\ell\ell}]_{1221}$, $[O_{ee}]_{1122}$, $[O_{\ell e}]_{1122}$, and $[O_{\ell e}]_{2211}$. 
In the limit of vanishing fermion masses, their effect on the forward ($\sigma_F$) and backward  ($\sigma_B$) $e^- e^+ \rightarrow \mu^- \mu^+$ cross sections  is given by  
\bea 
\label{eq:LEP2_sigmaFB}
\hspace{-1cm}
\delta \left (\sigma_F + \sigma_B \right )  &= &  {1 \over 24 \pi v^2} \left \{ 
 e^2  \left (  [c_{\ell \ell}]_{1122} + [c_{\ell \ell}]_{1221}  + [c_{e e}]_{1122} +  [c_{\ell e}]_{1122} +  [c_{\ell e}]_{2211}   \right )
 \right .  \nnl  &+ & \left. 
 {s  (g_L^2 + g_Y^2)   \over s- m_Z^2 } \left [  (g^{Ze}_{L,\rm SM})^2 \left ([c_{\ell \ell}]_{1122} + [c_{\ell \ell}]_{1221} \right )  +    (g^{Ze}_{R,\rm SM})^2 [c_{e e}]_{1122} 
 + g^{Ze}_{L,\rm SM}  g^{Ze}_{R,\rm SM} \left (  [c_{\ell e}]_{1122}  +  [c_{\ell e}]_{2211}  \right )
 \right ]  \right \}, 
 \nnl 
 \hspace{-1cm}
 \delta \left (\sigma_F -  \sigma_B \right )  &= &  {1 \over 32 \pi v^2} \left \{ 
  e^2   \left (  [c_{\ell \ell}]_{1122} + [c_{\ell \ell}]_{1221}  + [c_{e e}]_{1122} -   [c_{\ell e}]_{1122}  -  [c_{\ell e}]_{2211}   \right )
   \right .  \nnl  &+ & \left. 
 {s (g_L^2 + g_Y^2)  \over s- m_Z^2 } \left [  (g^{Ze}_{L,\rm SM})^2 \left ([c_{\ell \ell}]_{1122} + [c_{\ell \ell}]_{1221} \right )   +    (g^{Ze}_{R,\rm SM})^2 [c_{e e}]_{1122}  
 - g^{Ze}_{L,\rm SM}  g^{Ze}_{R,\rm SM} \left (  [c_{\ell e}]_{1122}  +  [c_{\ell e}]_{2211}  \right )
 \right ]  \right \}, 
 \nnl 
\eea 
where $g_{L,\rm SM}^{Ze}=-\frac{1}{2}+s^2_\theta $, $g_{R,\rm SM}^{Ze}=s^2_\theta$ are the couplings of the Z to left- and right-handed electrons. 
The effect of  the vertex corrections  $\delta g^{Ze}_L$, $\delta g^{Ze}_R$, $\delta g^{Z\mu}_L$, and $\delta g^{Z\mu}_R$  is also taken into account in the fit, but is not displayed here.  
The operator $[O_{\ell e}]_{1221}$  does not interfere with the SM due to the different helicity structure; thus it enters only at the quadratic ( $\cO(\Lambda^{-4})$) level and is neglected in this analysis. 

One observes that  measurements of the total cross section and asymmetry in $e^- e^+ \rightarrow \mu^- \mu^+$ in principle can constrain 3 linear combinations of the 5 four-lepton operators that enter in \eref{LEP2_sigmaFB}.   
 $[O_{\ell\ell}]_{1122}$ and $[O_{\ell\ell}]_{1221}$ are indistinguishable for this process because their parts involving charged leptons are related by a Fierz transformation.
$[O_{\ell e}]_{1122}$ and $[O_{\ell e}]_{2211}$ are also indistinguishable in this process, which can be traced to lepton flavor universality of the SM couplings. 
Accidentally, the LEP-2 observables depend very weakly on the  combination $[O_{\ell \ell}]_{1122} +[O_{\ell \ell}]_{1221} - [O_{ee}]_{1122}$  due to the fact that, numerically,  $(g_{L,\rm SM}^{Ze})^2 \approx (g_{R,\rm SM}^{Ze})^2$. \par

\medskip

We move to the process $e^- e^+ \rightarrow e^- e^+$ (Bhabha scattering).  
In  Ref.~\cite{Schael:2013ita}, LEP-2 quotes the differential cross sections for the scattering angle $\cos \theta$ in the interval $[-0.9,0.9]$, and the center-of-mass energies from 189~GeV to 207~GeV. 
Bhabha scattering is affected by the three four-leptons operators 
$[O_{\ell\ell}]_{1111}$, $[O_{ee}]_{1111}$ and $[O_{\ell e}]_{1111}$. 
In the limit of vanishing fermion masses their effect on the differential cross section is given by 
\begin{eqnarray}
\label{eq:lep2_e}
\delta \frac{d\sigma}{d\cos\theta} & = \frac{1}{8 \pi s}\frac{1}{v^2} & 
\left\lbrace u^2 \left[ e^2 ([c_{\ell \ell}]_{1111}+[c_{ee}]_{1111}) 
\left(\frac{1}{s}+\frac{1}{t} \right)\right. \right. \nonumber\\
& & + \left. (g_L^2+g_Y^2)\left(\left(g_{L,\rm SM}^{Ze}\right)^2 [c_{\ell\ell}]_{1111} +\left(g_{R,\rm SM}^{Ze}\right)^2 [c_{ee}]_{1111} \right)\left( \frac{1}{s-m_Z^2}+\frac{1}{t-m_Z^2} \right) \right] \nnl 
& & + \left. t^2\left[ [c_{\ell e}]_{1111} \frac{e^2}{s} 
+ [c_{\ell e}]_{1111} \frac{(g_L^2+g_Y^2)g_{L,\rm SM}^{Ze} g_{R,\rm SM}^{Ze}}{s-m_Z^2}\right]
\right. \nnl & &  \left.  
+s^2\left[ [c_{\ell e}]_{1111} \frac{e^2}{t} + [c_{\ell e}]_{1111}
 \frac{(g_L^2+g_Y^2)g_{L,\rm SM}^{Ze} g_{R,\rm SM}^{Ze}}{t-m_Z^2}\right] \right\rbrace,  
\end{eqnarray} 
where  $t = -\frac{s}{2}(1-\cos \theta)$ and $u = -\frac{s}{2}(1+\cos \theta)$.
Again, the dependence on the vertex  corrections $\delta g^{Ze}_L$, $\delta g^{Ze}_R$ is taken into account in our analysis but not displayed here. 
In principle,  Bhabha scattering at LEP-2 constrains independently all 3 four-electron operators, 
but again an approximate flat direction along the direction $[O_{\ell\ell}]_{1111} -  [O_{ee}]_{1111}$ arises due to the numerical accident $(g_{L,\rm SM}^{Ze})^2 \approx (g_{R,\rm SM}^{Ze})^2$.

\subsection{Low-energy neutrino scattering}

Interactions of SM leptons can be probed by neutrino scattering on electrons.  
We focus on processes with muon neutrinos: $\nu_\mu \; e^- \rightarrow \nu_\mu \; e^-$,  and muon anti-neutrinos: $\overline{\nu}_\mu \; e^- \rightarrow \overline{\nu}_\mu \; e^-$, which were studied at center-of-mass energies far below the Z-pole by  the CHARM \cite{Dorenbosch:1988is}, CHARM-II \cite{Vilain:1994qy}, and BNL-734   \cite{Ahrens:1990fp} experiments.  
The results are usually presented as constraints on the vector ($g_V$) and axial ($g_A$) coupling strength of the Z boson to electrons: 
\begin{equation}
\begin{tabular}{|l|c|c|c|}
   \hline
    Experiment & Ref. & $g_V$ & $g_A$ \\
    \hline
  CHARM-II & \cite{Vilain:1994qy} & $-0.035\pm0.017$  &  $-0.503\pm0.0017$\\
   \hline
  CHARM & \cite{Dorenbosch:1988is} & $-0.06\pm 0.07$  & $-0.54\pm0.07$\\
   \hline
  BNL-E734 & \cite{Ahrens:1990fp} & $-0.107\pm 0.045$  & $-0.514\pm0.036$\\
   \hline
 \end{tabular}
\end{equation}
where the SM predicts  $g_V = -0.0396$, $g_A = -0.5064$ \cite{Erler:2013xha}. 

In the presence of $D$=6 operators, the scattering cross sections measured in these experiments are sensitive not only to the Z boson couplings but also to the four-leptons operators involving the 2nd generation doublet: $[O_{\ell\ell}]_{1122}$ and $[O_{\ell e}]_{2211}$. 
Nevertheless,  at energies below the Z-pole, the measurements of $g_V$ and $g_A$ can be easily recast as constraints on the parameters in our framework. 
At the linear level, the vector and axial couplings  are {\em effectively} modified as 
\begin{eqnarray}
\delta g_V &=& \delta g^{Ze}_L+\delta g^{Ze}_R+\frac{3g_Y^2-g_L^2}{g_L^2+g_Y^2}\left(\delta g^{Z\mu}_L+\delta g^{W\mu}_L\right)- \frac{[c_{\ell\ell}]_{1122}+[c_{\ell e}]_{2211}}{2}, 
\nnl 
\delta g_A &=&\delta g^{Ze}_L-\delta g^{Ze}_R -\left(\delta g^{Z\mu}_L+\delta g^{W\mu}_L\right)- \frac{[c_{\ell\ell}]_{1122}-[c_{\ell e}]_{2211}}{2}. 
\end{eqnarray}
Notice that the dependence on the four-lepton operators is different than for the LEP-2 observables discussed in the previous subsection. 
Therefore, low-energy neutrino scattering provides us with complementary information that will allow us to constrain additional directions in the space of $D$=6 Wilson coefficients.  

Experimental results on low-energy scattering of {\em electron} neutrinos \cite{Auerbach:2001wg} and anti-neutrinos \cite{Abe:2012gx} on electrons are also available. 
These probe the 4-electron operators $[O_{\ell \ell}]_{1111}$ and $[O_{\ell e}]_{1111}$.  
However, the current experimental accuracy is worse than for muon neutrinos scattering, 
and including this additional input would not affect the global fit in an appreciable way.

\subsection{Parity violating electron scattering}

The SLAC E158 experiment made a precise measurement of parity-violating asymmetry in \moller scattering $e^- e^- \rightarrow e^- e^-$  \cite{Anthony:2005pm}.
The asymmetry is defined as $A_{PV}=(\sigma_R-\sigma_L)/(\sigma_R+\sigma_L)$ where $\sigma_{L(R)}$ is the cross-section for incident left- (right-) handed electrons. The E158 experiment used a polarized electron beam of energy $E\approx50$ GeV against an electron target at rest which corresponds to a center-of-mass energy of $\sqrt s \approx \sqrt{2 m_e E} \approx 0.2$ GeV, far below the $Z$ pole. 
The results are presented as a measurement of the weak mixing angle at low energies:  
\begin{equation}
s^2_\theta(Q^2=0.026 \gev^2)=0.2397\pm 0.0013, 
\end{equation}
where the SM predicts  $s^2_\theta(Q^2=0.026 \gev^2)=0.2381\pm 0.0006$ \cite{Czarnecki:1995fw}.

$A_{PV}$ in  \moller scattering is sensitive to the four-electron operators $[O_{ee}]_{1111}$ and $[O_{\ell\ell}]_{1111}$ ($[O_{\ell e}]_{1111}$ cancels out in $\sigma_R-\sigma_L$). 
At the linear order in the EFT parameters and leading order in $s/m_Z^2$, 
the effect of these operators and the vertex corrections  can be effectively represented as a shift of the measured weak mixing angle:  
\begin{equation}
\delta s^2_\theta=
2(g_{R,SM}^{Ze}  \delta g_R^{Ze} -g_{L,SM}^{Ze} \delta g_L^{Ze} )
-\frac{1}{4}([c_{ee}]_{1111}-[c_{\ell\ell}]_{1111})
\end{equation}
Although \moller scattering probes the same 4-electron operators as LEP-2, c.f. \eref{lep2_e},  
its importance rests in the  sensitivity to the combination that is accidentally very weakly constrained by unpolarized electron scattering in LEP-2.

\subsection{Tau and muon decays}
\label{sec:tmd}

The leptonic tau decays $\tau^- \rightarrow e^- \nu_\tau \bar\nu_e$,  $\tau^- \rightarrow \mu^- \nu_\tau \bar\nu_\mu$, and the conjugates provide additional information on 4-lepton operators involving $\tau$. 
In particular, the provide the only constraint we are aware of on lepton-flavor conserving 4-lepton operators with muons and taus. 
The  decays can be described by the following effective Lagrangian:
\beq
\mathcal{L}= -\frac{4G_{\tau f}}{\sqrt{2}}(\bar{\nu}_\tau \bar \sigma_\rho \tau)(\bar f \bar \sigma_\rho \nu_f)
+ \hc , 
\eeq
where $f = e,\mu$. 
At the linear level,  the relative strength of the Fermi constant measured  in the tau decays normalized to that measured  in the muon decay is affected by the vertex corrections and four-lepton operators as 
\bea
A_e &\equiv  & \frac{G_{\tau e}^2}{G_F^2}=1+ 2 \delta g^{W \tau}_L+2 \delta g^{W e}_L -4\delta m - [c_{\ell \ell}]_{1331}, 
\nnl 
A_\mu &\equiv  & \frac{G_{\tau \mu}^2}{G_F^2}=1+ 2 \delta g^{W \tau}_L+2 \delta g^{W \mu}_L -4\delta m - [c_{\ell \ell}]_{2332}, 
\eea 
where the W mass corrections $\delta m$ can be expressed by other EFT parameters, c.f. \eref{dm}.  
The experimental  values quoted by the PDG are \cite{Agashe:2014kda} 
\bea
A_e & = & 1.0029 \pm 0.0046, 
\nonumber\\
A_\mu &= & 0.981 \pm 0.018, 
\eea
and the SM prediction is  $A_f = 1$. 

For the  muon decay, $\mu^- \rightarrow e^- \nu_\mu \bar\nu_e$ and the conjugate, the total rate defines  the SM input parameter $v$ and by itself it does not probe new physics. 
However, additional information can be extracted from differential distributions in (polarized) muon decay. 
Customarily, these measurements are presented in the language of {\em Michel parameters} \cite{Bouchiat:1957zz}. 
From the EFT perspective the most interesting are the so-called $\eta$ and $\beta'/A$ parameters, 
because they are the only ones that may receive contributions at $\cO(1/\Lambda^2)$ \cite{Gonzalez-Alonso:2014bga,martinthesis}: 
\beq
\eta = {{\rm Re} [c_{\ell e}]_{1221}  \over 2}, 
\qquad 
\beta'/A = - { {\rm Im} [c_{\ell e}]_{1221}\over 4}. 
\eeq 
These parameters have been measured in an experiment in the PSI \cite{Danneberg:2005xv}: 
\beq
\eta = -0.0021 \pm 0.0071, 
\qquad 
\beta'/A =  -0.0013 \pm 0.0036. 
\eeq 
Analogous limits from tau decays are much weaker.

\section{General Fit}
\label{sec:fit}

We now do a global fit to all the data discussed above so as to simultaneously constrain 
$D$=6 operators in the EFT Lagrangian that give rise  to  leptonic vertex corrections and 4-lepton interactions.
Previously, constraints on 4-lepton (and other 4-fermion) operators were obtained in Refs.~\cite{Han:2004az,Han:2005pr} and recently updated in Ref.~\cite{Berthier:2015gja}, assuming the Wilson coefficients are the same for all 3 fermion generations.  
The novel aspect of our analysis is that we allow for a completely general  flavor structure of the $D$=6 operators. 

We combine the following  experimental inputs discussed in \sref{exp}:  
\bi 
\item Z boson production and decay in LEP-1 and leptonic W decays in LEP-2, 
\item W mass measurement, 
\item Two-lepton production in LEP-2,  
\item Muon-neutrino scattering on electrons,
\item Parity violation in low-energy \moller scattering, 
\item $G_F$ measurements in $\tau$ decays.  
\ei 

We consider the EFT Lagrangian with operators up to $D$=6, neglecting possible contributions of $D$=8 operators.\footnote{
In the EFT expansion,  Wilson coefficients of $D$=8 operators are suppressed by another factor of $v^2/\Lambda^2$ compared to those of $D$=6 operators. Thus, they are generically subleading  when the EFT approach is valid, that is when the new physics scale $\Lambda$ is greater than the electroweak scale.    
Exceptions to that rule could occur if symmetries or fine-tuning in the UV theory lead to a suppression of some $D$=6 (but not the corresponding $D=8$ or higher) Wilson coefficients in the low-energy EFT. 
Our results are not valid in such situations;  see Ref.~\cite{Berthier:2015gja} for a discussion relevant to these cases.  
}
Consistently, in our analysis we only include corrections to observables that are linear in Wilson coefficients of $D$=6 operators. 
These are formally $\cO(v^2/\Lambda^2)$ in the EFT counting, and come from interference between tree-level SM and $D$=6 contributions to the relevant amplitudes. 
We also ignore loop-suppressed effects proportional to $D$=6 Wilson coefficients.   
We use the experimental results, the SM predictions, and the analytic expression for $D$=6 contributions discussed in \sref{exp} to construct a global Gaussian likelihood in the space of the relevant Wilson coefficients.  
With this procedure, we get the following global constraints:
\beq
\label{eq:generalfit}
\bvec 
\delta g^{We}_L \\ 
\delta g^{W\mu}_L \\ 
\delta g^{W\tau}_L \\ 
\delta g^{Ze}_L \\ 
\delta g^{Z\mu}_L \\ 
\delta g^{Z\tau}_L \\ 
\delta g^{Ze}_R  \\ 
\delta g^{Z\mu}_R \\ 
\delta g^{Z\tau}_R \\ 
\, [c_{\ell \ell}]_{1111} \\ 
 \, [c_{\ell e}]_{1111} \\     
\,  [c_{e e}]_{1111}   \\    
\, [c_{\ell \ell}]_{1221} \\ 
\, [c_{\ell \ell}]_{1122} \\ 
 \, [c_{\ell e}]_{1122} \\   
 \, [c_{\ell e}]_{2211} \\    
\,  [c_{e e}]_{1122}  \\     
\, [c_{\ell \ell}]_{1331} \\ 
\, [c_{\ell \ell}]_{1133} \\ 
 \, [c_{\ell e}]_{1133}+  [c_{\ell e}]_{3311} \\   
\,  [c_{e e}]_{1133}     \\ 
\, [c_{\ell \ell}]_{2332} \\ 
\evec    
= 
\bvec
 -1.00 \pm 0.64 \\   -1.36  \pm 0.59 \\  1.95 \pm 0.79 \\
 -0.027 \pm 0.028 \\   0.01  \pm 0.11 \\  0.016  \pm 0.058 \\  
 -0.037 \pm 0.027 \\   0.00  \pm 0.13 \\  0.039  \pm 0.062 \\ 
0.99 \pm 0.39 \\  -0.23  \pm  0.22 \\ 0.23 \pm 0.39 \\ 
-4.8 \pm 1.6 \\  2.0  \pm  2.3 \\ 0.9 \pm 2.3  \\-0.8 \pm 2.2 \\  2.8 \pm 2.8 \\ 
1.5 \pm 1.3 \\ 
140  \pm  170 \\
-0.55  \pm 0.64  \\ 
-150 \pm 180 \\ 
3.0 \pm 2.3 
\evec 
 \times 10^{-2}, 
\eeq 
with the correlation matrix written down in \eref{fullmonty_rho}.  

A few general comments are in order: 
\bi
\item 
In the global fit, the constraints on the leptonic vertex corrections are the same as the ones in \eref{dgl} determined from on-shell Z and W data. 
The additional experimental input considered in this analysis constrains 4-lepton operators without affecting the limits on  the vertex corrections in an appreciable way. 
Nevertheless, the correlations between vertex corrections and 4-lepton operators are non-negligible in some cases, as can be observed in   \eref{fullmonty_rho}. 
\item Not all 4-lepton operators can be constrained by the current data.
In particular, we are not aware of any experiments probing four-muon or four-tau interactions. 
On the other hand, most of the Wilson coefficients of 4-lepton operators involving electrons are constrained, in a model-independent way,  at a percent level accuracy.  
\item In \eref{generalfit}, the limits on the electron-tau 4-fermion operators 
$[O_{\ell \ell}]_{1133}$ and $[O_{e e}]_{1133}$ are very weak.  
Actually, the combination $[O_{\ell \ell}]_{1133} +   [O_{e e}]_{1133}$ is constrained at a percent level. 
However,  $\tau$-pair production in LEP-2 is  accidentally insensitive to   $[O_{\ell \ell}]_{1133} -  [O_{e e}]_{1133}$, as discussed  in \sref{lep2}, 
and this is reflected in our fit by the $-1$ value  of the corresponding correlation coefficient.  
Moreover, only the sum $[O_{\ell e}]_{1133} + [O_{\ell e}]_{3311}$ can be probed in LEP-2.  
We note that both flat directions would be absent if polarization of the colliding electrons was known.    
Measurements with polarized $e^\pm$ beams in future linear colliders \cite{Baer:2013cma,Koratzinos:2013ncw} will provide additional information that will break these degeneracies and greatly improve model-independent constraints on electron-tau 4-fermion operators. 
For electron-muon operators the corresponding  flat direction is absent in \eref{generalfit} thanks to  including the experimental input  from muon neutrino scattering on electrons.  
From this point of view, it would be extremely interesting to sturdy tau neutrino scattering on electrons, although we are not aware of any realistic experimental plans in this direction.
For  4-electron operators,  the direction $[O_{\ell \ell}]_{1111} -  [O_{e e}]_{1111}$ is also practically unconstrained by LEP-2,  but in this case the degeneracy is lifted thanks to  parity violating \moller scattering.
\item 
In \eref{generalfit} we do not give any constraints on $[c_{\ell e}]_{IJJI}$ with $I \neq J$.
That is because, in the limit $m_{e_I}  = 0$,  the corresponding operators do not interfere with the SM, thus they contribute to  the observables at $\cO(\Lambda^{-4})$ and are neglected.  
However, as discussed in \sref{tmd}, they contribute  at $\cO(\Lambda^{-2})$ to the Michel parameters 
$\eta$ and $\beta'/A$ in $e_I \to e_J \nu \nu$ decays (which are in fact defined only for $m_{e_I}  > 0$).  
The experimental  limits on the Michel parameters measured in muon decays  translate to 
\beq
\label{eq:cle1221}
\mathrm{Re}([c_{\ell e}]_{1221}) = (-0.4 \pm 1.4 ) \times 10^{-2}, 
\qquad 
\mathrm{Im}([c_{\ell e}]_{1221}) = (0.5 \pm 1.4) \times 10^{-2}.
\eeq 
One can also constrain the analogous operators with tau leptons. 
Translating the constraints on  the form factor $g^S_{RR}$ in leptonic tau decays \cite{Agashe:2014kda} one obtains:  
$\mathrm{Re}([c_{\ell e}]_{2332}) = 0.19 \pm 0.15$, 
and $|[c_{\ell e}]_{1331}| < 0.70$, $|[c_{\ell e}]_{2332}| < 0.72$  at  95\% confidence level (CL). 
Stronger limits may arise via  1-loop contributions of these operators to anomalous electric and magnetic moments \cite{Crivellin:2013hpa}, however in this case the limits concern in fact for a linear combination of $[c_{\ell e}]_{IJJI}$ and the Wilson coefficients of $D$=6 dipole operators in the effective Lagrangian.
\item 
Combining the results from this paper with the ones in  Ref.~\cite{Efrati:2015eaa} one could also perform a global analysis of leptonic parameters together with quark vertex corrections. 
However, correlations between 4-lepton operators and quark vertex corrections are small:  we find that the correlation coefficients are typically of order $0.01$, and the largest is $0.07$.   
All in all, the constraints on the quark vertex corrections  and their correlations with the leptonic vertex  corrections quoted in \cite{Efrati:2015eaa}  are not affected by the combination. 
\ei 

Using the results in \eref{generalfit} and the correlation matrix $\rho$ one can reconstruct 
the complete Gaussian likelihood function in the space of leptonic vertex corrections and four-lepton operators: 
\beq
\label{eq:chi2}
\chi^2=   \sum_{ij}[x - x_0]_{i} \sigma^{-2}_{ij}   [x - \delta x_0]_{j},
\eeq
where $\sigma^{-2}_{ij} =  [[\Delta x]_i \rho_{ij} [ \Delta x]_j]^{-1}$, 
$\vec x$ is a 14-dimensional vector collecting the $\delta g$'s and  $c$'s as in \eref{generalfit},   
 and $\vec x_0$, $\Delta \vec x$ are the corresponding central values and $1~\sigma$ errors.  
 In specific extensions of the SM only a subset of the general EFT parameters will be generated. 
In such a case, constraints on the model parameters can be obtained by restricting the full likelihood to the smaller subspace, and then minimizing the restricted likelihood. 
In the next section we perform this procedure for a handful of scenarios beyond the SM that affect only leptonic observables.

\section{Leptophilic Models}
\label{sec:models}

\subsection{One by one}

\begin{figure}[htb]
\bc
\includegraphics[width=0.95 \textwidth]{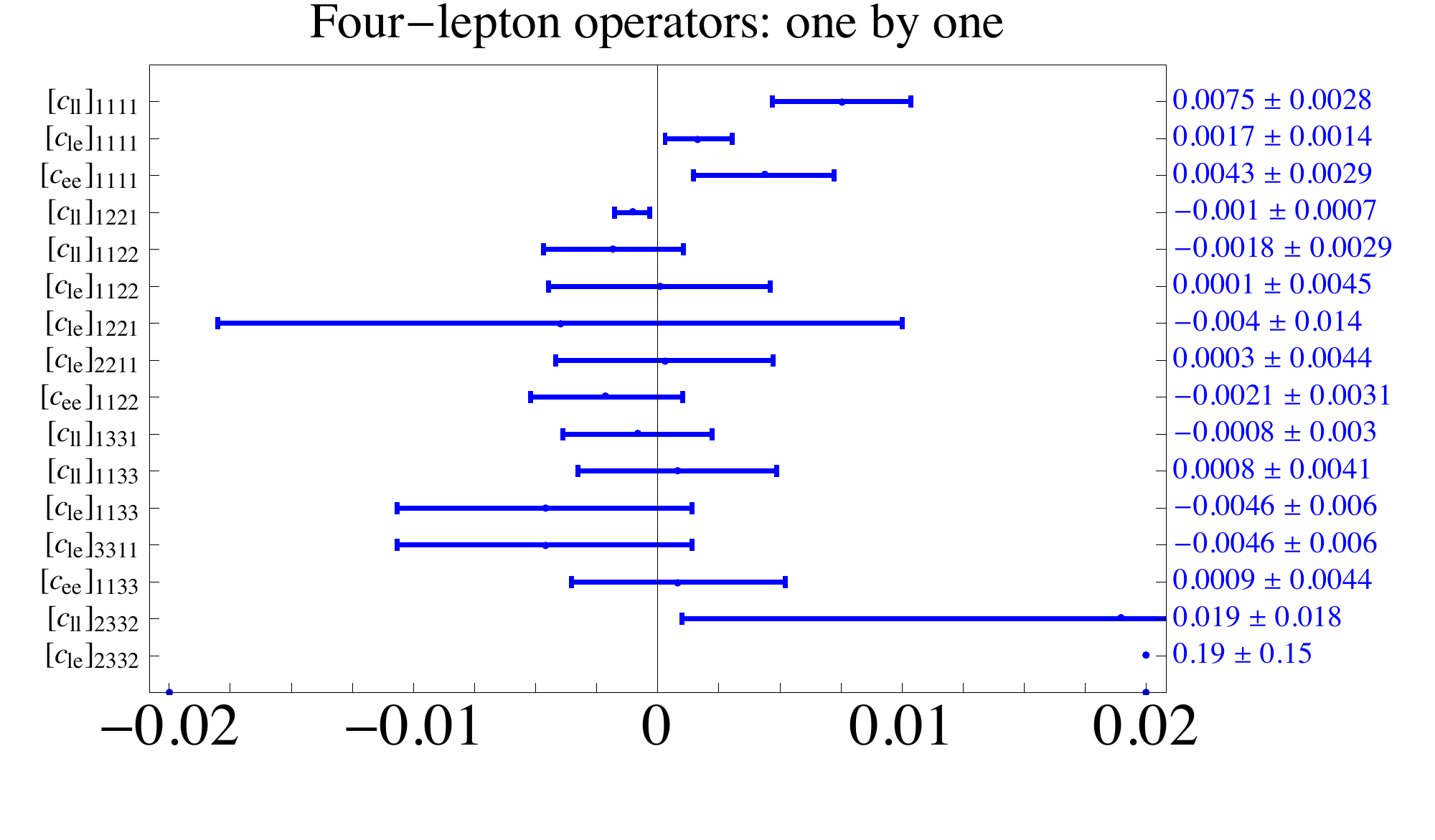}
\ec 
\caption{
 Summary of the 68\% CL intervals for the Wilson coefficients of 4-lepton operators, assuming that only 1 such operator is present at a time and that  leptonic vertex corrections are absent.  
}
\label{fig:onebyone_c}
\end{figure}  

Before we attack specific models, we first discuss a general scenario where only one four-lepton operator and no vertex corrections is generated by new physics. 
Setting all but one Wilson coefficient  to zero in the likelihood in \eref{chi2}, 
and then minimizing the resulting function we obtain the constraints summarized in \fref{onebyone_c}.
The strongest constraint, at a per-mille level, is the one on $[c_{\ell \ell}]_{1221}$. 
The reason is that the corresponding operator affects the measurement of the Fermi constant $G_F$ in muon decays, and this way, unlike other 4-fermion operators, it affects the electroweak precision observables very accurately measured  in LEP-1.
The constraints on the remaining operators containing electrons  are dominated by lepton pair production in LEP-2 and are somewhat weaker. 
Finally, the muon-tau four-lepton operators are only weakly constrained by tau decays. 

We also visualize these constraint in terms of the new physics scale probed by each operator. 
To this end, we write $c_i = \pm g_*^2/\Lambda_i^2$, where  $\Lambda_i$ can be interpreted as the mass scale of new particles and $g_*$ their coupling strength.   
Then we  derive the 95\% CL lower limit on the ratio $\Lambda_i/g_*$. 
In general, the limit depends on the sign of the Wilson coefficient, and  for our presentation 
 we always choose the lower one of the two possibilities. 
The results are shown in  \fref{onebyone_lambda}.  
Current data allow one to probe new particles with masses  up to 5 TeV if they are coupled to the SM with order one strength.

\begin{figure}[htb]
\bc
\includegraphics[width=0.95 \textwidth]{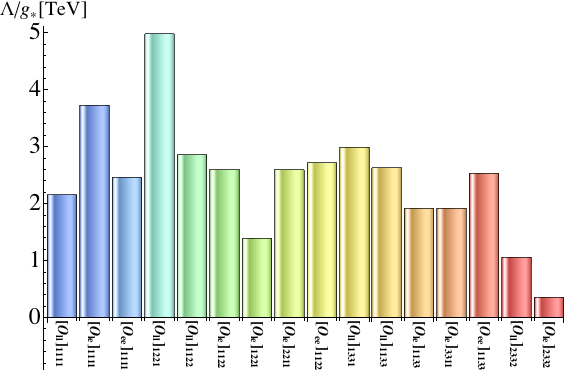}
\ec 
\caption{
95\% CL lower limits on the scale suppressing 4-lepton operators in the EFT Lagrangian, assuming that only 1 such operator is present at a time and that  leptonic vertex corrections are absent.  
}
\label{fig:onebyone_lambda}
\end{figure} 

\subsection{Z prime}

We now consider a model with a new neutral vector boson $V_\mu$ of mass $m_V$ coupled to leptons as 
\beq
\cL \supset V_\mu \left ( 
\kappa_{L,I} \bar \ell_I \bar  \sigma_\mu \ell_I  + \kappa_{R,I} e^c_I   \sigma_\mu \bar e^c_I . 
\right )
\eeq
We assume the vector does not mix with the Z-boson, and does not couple to quarks. 
In such a case, it can be constrained neither by Z-pole observables nor by LHC dilepton resonance searches. 
However, it is constrained by the off-Z-pole observables discussed here. 
Integrating out the vector we get an effective theory with the Wilson coefficient of 4-lepton operators: 
\bea
\, [c_{\ell \ell}]_{IIJJ} &= & -  \kappa_{L,I} \kappa_{L,J}{v^2 \over m_V^2},
\nnl 
\, [c_{ee}]_{IIJJ} &= &  - \kappa_{R,I} \kappa_{R,J}{v^2 \over m_V^2},
\nnl 
\, [c_{\ell e}]_{IIJJ} &= & -  \kappa_{L,I} \kappa_{R,J} {v^2 \over m_V^2}. 
\eea
Plugging these expressions in the general likelihood in \eref{chi2} we obtain the likelihood as a  function of  $\kappa/m_V$.
In \fref{zprime} we show examples of this likelihood  for 2 scenarios: 
one where the vector couples to electrons only, and another one where the vector couples universally to all leptons,  $\kappa_{L/R,I}=  \kappa_{L/R}$. 
In both cases, we find that  $\kappa/m_V \lesssim 0.1$-$0.3/$TeV, depending on the ratio of the left- and right-handed couplings. 
We can also observe that the vector-like  couplings, $\kappa_L \approx \kappa_R$, are more strongly constrained than the axial ones,    $\kappa_L \approx  - \kappa_R$. 

\begin{figure}[htb]
\bc 
\includegraphics[width=0.45 \textwidth]{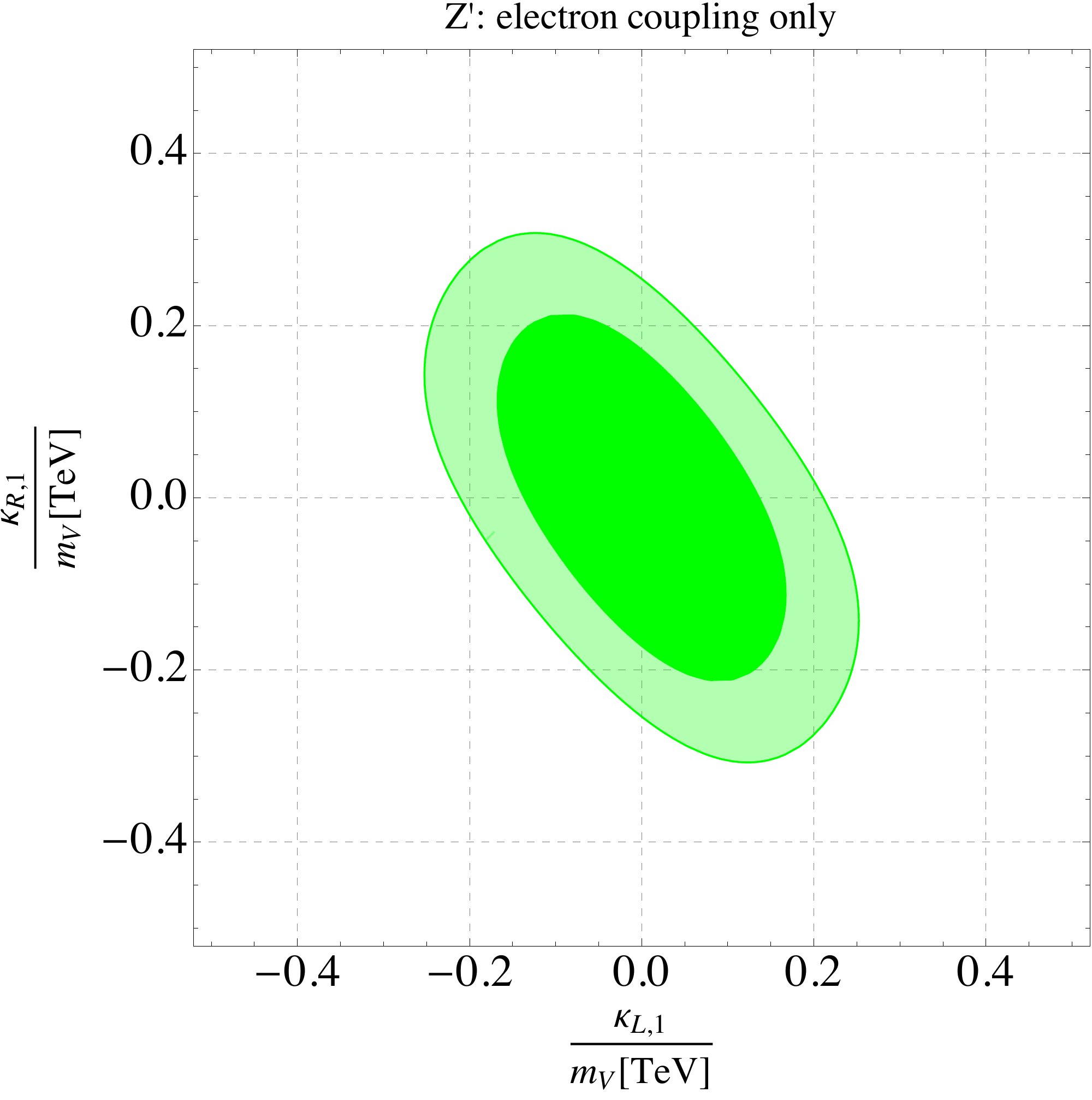}
\quad 
\includegraphics[width=0.45 \textwidth]{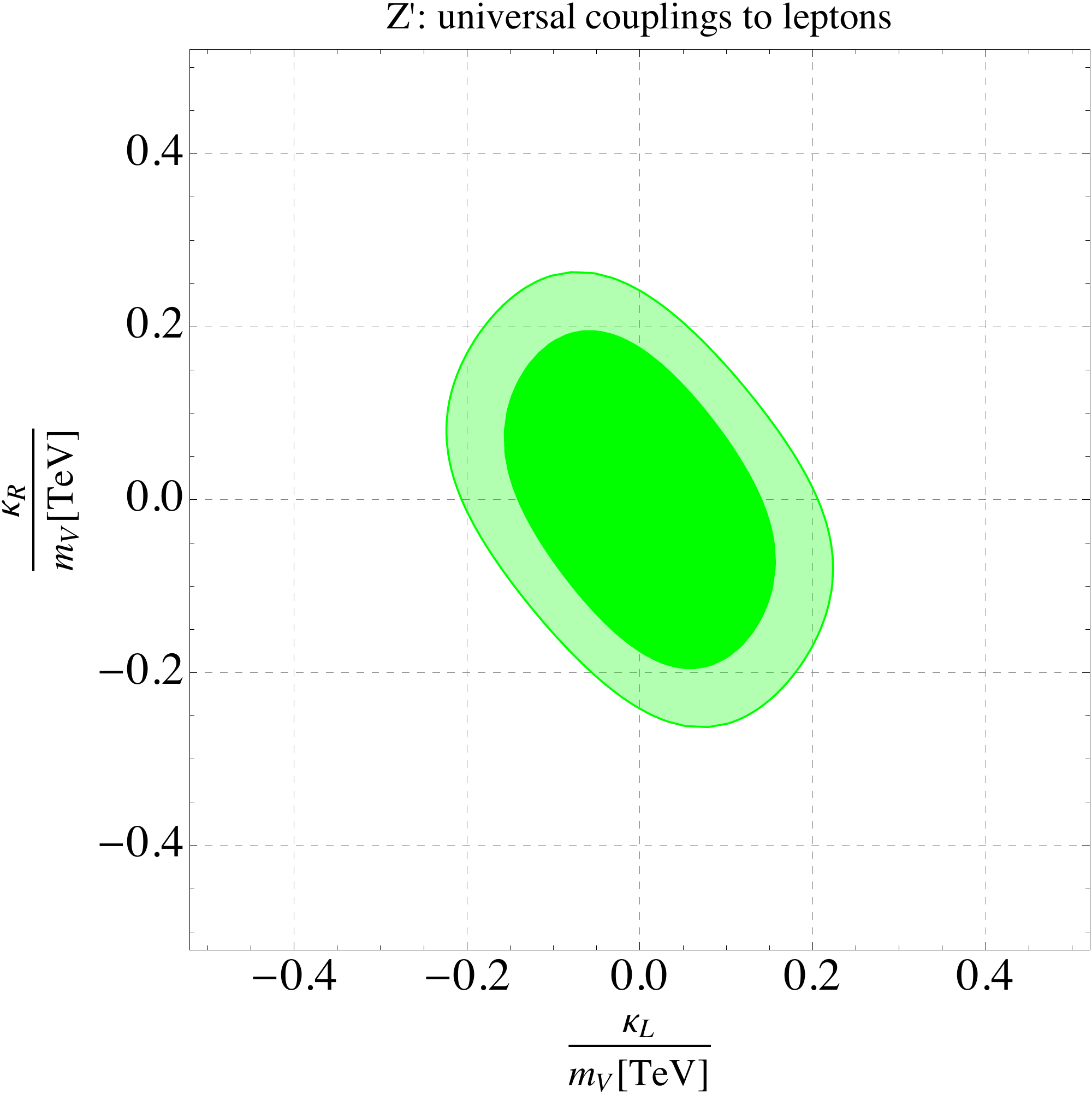}
\ec 
\caption{
{\em Left:}   68~\% CL (darker green) and  95~\% CL (lighter green) regions  for  the ratio of the couplings over mass for the leptophilic Z' vector boson coupled to electrons only.    
{\em Right:} The same for Z' coupled universally to all leptons.  
}
\label{fig:zprime}
\end{figure}

\subsection{Vector Triplet}

Consider a model with a new triplet of  vector bosons $V_\mu^i$ of mass $m_V$ coupled to left-handed leptons as 
\beq
\cL \supset \kappa_{I}  V_\mu^i  \bar \ell_I \bar  \sigma_\mu {\sigma^i \over 2} \ell_I  . 
\eeq
Integrating out the triplet we get the following 4-lepton operators in the EFT:
\beq
\cL_{\rm eff} \supset - {\kappa_{I} \kappa_{J} \over 8 m_V^2}   (\bar \ell_I \bar  \sigma_\mu \sigma^i \ell_I )  
  (\bar \ell_J \bar  \sigma_\mu \sigma^i \ell_J  )   . 
\eeq   
These operators were not  introduced previously.  
The reason is that they are related to other 4-lepton operators in \tref{4l} via Fierz transformations. 
Using, the identity  
\beq
  (\bar \ell_I \bar  \sigma_\mu \sigma^i \ell_I  )    (\bar \ell_J \bar  \sigma_\mu \sigma^i \ell_J)   = 
2   (\bar \ell_I \bar  \sigma_\mu \ell_J )    (\bar \ell_J \bar  \sigma_\mu  \ell_I) 
-  (\bar \ell_I \bar  \sigma_\mu \ell_I )    (\bar \ell_J \bar  \sigma_\mu  \ell_J), 
\eeq 
one identifies  the Wilson coefficients in the low-energy EFT as 
\bea
\, [c_{\ell \ell}]_{IIII} &= & -  \kappa_{I}^2 {v^2 \over 4 m_V^2},
\nnl 
\, [c_{\ell \ell}]_{IJJI} &= & -  \kappa_{I}  \kappa_{J} {v^2 \over 2 m_V^2}, \quad I < J, 
\nnl 
\, [c_{\ell \ell}]_{IIJJ} &= &  \kappa_{I}  \kappa_{J} {v^2 \over 4 m_V^2}, \quad I < J. 
\eea 
When the triplet couples to electrons only, $\kappa_1 \neq 0$ and $\kappa_{2,3}=0$, only one four-lepton operator $[O_{\ell \ell}]_{1111}$ is generated.  
Plugging the above expression in  the general likelihood in \eref{chi2} one obtains the following bound on the ratio of the vector mass and coupling: 
\beq
\label{eq:triplet}
{m_V \over \kappa_1 } \geq 2.9~{\rm TeV}, \qquad @ \, 95\% \, {\rm CL}.
\eeq 
The limit is stronger than what might be inferred from the one-by-one limits plotted in \fref{onebyone_lambda}, because the model predicts the negative sign of the Wilson coefficient $[c_{\ell \ell}]_{1111}$, for which the experimental constraints are stronger than for the positive one. 
On the other hand, when the triplet couples with the same strength to all leptons, $\kappa_I =  \kappa$, 
we find a slightly stronger bound:  
\beq
\label{eq:triplet_uni}
{m_V \over \kappa } \geq 3.7~{\rm TeV}, \qquad @ \, 95\% \, {\rm CL}.
\eeq

\subsection{Inert Higgs}

The last example we study is  a model with a scalar  $S$ of mass $m_S$
 transforming, much like the SM Higgs,  as $2_{1/2}$ under $SU(2)_L \times U(1)_Y$, and coupled to leptons as 
\beq
\cL \supset - S^\dagger  Y_{I} \ell_I  e^c_I + \hc   .
\eeq
We assume that $S$ does not get a VEV. 
Integrating out the scalar we get the following 4-lepton operators in the EFT:
\beq
\cL_{\rm eff} \supset {Y_{I}^* Y_{J} \over m_S^2}  (\bar \ell_I  \bar e^c_I )   (\ell_J  e^c_J).   
\eeq   
Once again these operators do not appear in \tref{4l}, but  using the Fierz transformation, 
$\bar \sigma^{\dot \alpha \alpha}_\mu  \sigma_{\beta \dot  \beta}^\mu = 2 \delta^\alpha_\beta \delta^{\dot \alpha}_{\dot \beta}$  we can rewrite them as 
\beq
\cL_{\rm eff} \supset {Y_{I}^* Y_{J} \over 2 m_S^2} 
(\bar \ell_I \bar \sigma_\mu  \ell_J)   (e^c_J  \sigma_\mu \bar e^c_I ).   
\eeq   
This way, we identify the Wilson coefficients of the 4-lepton operators induced in the EFT by integrating out  the scalar $S$:
\beq
[c_{\ell e}]_{IJJI}  = Y_{I}^* Y_{J} {v^2  \over 2 m_S^2} . 
\eeq 
When the scalar couples to electrons only, $Y_1 \neq 0$ and $Y_2 = Y_3 = 0$, 
its mass over coupling is constrained as  
\beq
\label{eq:higgs}
{m_S \over |Y_1| } \geq 2.6~{\rm TeV}, \qquad @ \, 95\% \, {\rm CL}.
\eeq 
When the scalar couples to all 3 generations of leptons  then the  constraints on
the Michel parameters discussed around \eref{cle1221} can be relevant. 
If the couplings $Y_I$ are the same for all 3 generations, $Y_I = Y$, then these constraints have a small effect, and the bound $m_S/|Y|$ is the same as in  \eref{higgs}. 
However, if   $Y_I $ are proportional to fermion's mass, $Y_I =  Y_3\, m_{e_I}/m_{\tau}$, then   
the constraint on $[c_{\ell e}]_{2332}$ from tau decays is the dominant one, 
leading to $m_S/|Y_3| > 60$~GeV at 95\% CL. 

\section{Conclusions}

In this paper we discussed constraints on 4-lepton $D$=6 operators in the EFT beyond the SM. 
For the first time, the analysis was performed without any assumptions about the flavor structure of the $D$=6 operators.  
We presented our results such that they can readily be recast as constraints on specific models beyond the SM  that, after integrating the new heavy particles, lead to  leptonic vertex corrections and 4-lepton operators in the low-energy EFT. 
Our results are particularly relevant in relation to models where lepton flavor universality is not preserved. 

We  find that  the typical current experimental  sensitivity to the scale suppressing  4-lepton operators is of order a few TeV.  
In the best case, one can probe 50 TeV particles provided they interact strongly with the SM leptons ($g_* \sim 4\pi$), and they generate the best constrained operator  $[O_{\ell \ell}]_{1221}$ (e.g, via exchange of an $SU(2)_L$ triplet of vector bosons).
Our analysis also reveals several blind spots where the current sensitivity is weaker, which  would be interesting targets for future experiments.
In particular, certain linear combinations of 4-lepton operators involving electrons and taus are very weakly constrained. 
This can be cured by future $e^+ e^-$ colliders once polarization information about the initial state is available.  
Moreover,  experimental information on 4-lepton operators involving muons and taus is currently very limited. 
Bounds on two such operators involving left-handed doublets could be improved by more precise measurements of the rate and differential distributions in $\tau \to \mu \nu \nu$ decays; 
probing the remaining operators  would be  one of the strong points of the physics program of a future  $\mu^+ \mu^-$ collider.

Including  the available experimental information about  $e^+ e^- \to$~jets in LEP-2  and $p p \to$~leptons and jets at the LHC, as well as about low energy scattering on nuclei one can generalize this analysis so as to also constrain 4-fermion operators involving quarks. 
This is left for a future publication.

\section*{Acknowledgements}
  
AF~is supported by the ERC Advanced Grant Higgs@LHC.

\appendix

\section{Relation to universal theories}
\label{app:uni}

In this paper we used the formalism where new physics effects in the $D$=6 EFT were represented by vertex corrections and 4-fermion operators.  
At the same time, the  quadratic terms of electroweak gauge bosons in the Lagrangian were assumed to be the same as in the SM, except for a correction to the W boson mass. 
That can always be achieved without loss of generality, via field redefinitions and integration by parts. 
On the other hand, in the literature, precision constraints on new physics are often expressed in the language of {\em oblique parameters}, such as the Peskin-Takeuchi $S$, $T$, $U$ parameters \cite{Peskin:1991sw}. 
These are, on the contrary, defined via corrections to kinetic terms of electroweak gauge bosons.   
In this appendix we discuss the relationship between the two formalisms.  

Oblique parameters fully characterize the new physics effects for the so-called {\em universal} theories \cite{Barbieri:2004qk,Wells:2015uba}.
The theory is universal  if one can recast it to a form where  new physics affects only propagators of the SM gauge bosons.  
Before introducing the oblique parameters, we first define the momentum expansion of the 2-point functions of electroweak gauge bosons:   
\beq
 \cM(V_{1,\mu} \to V_{2,\nu}) =   \eta_{\mu\nu} \left ( \Pi_{V_1 V_2}^{(0)} +   \Pi_{V_1V_2}^{(2)} p^2 +   \Pi_{V_1V_2}^{(4)} p^4 +   \dots \right )  
+ p_{\mu}p_{\nu}  \left ( \dots \right), 
\eeq
where $p$ is the 4-momentum of the incoming gauge boson. 
We are interested in corrections  $\delta \Pi_{V_1 V_2}^{(n)}$ with respect to the 2-point function in the SM. 
Two-point functions are not directly  measurable, but certain combinations of  $\delta \Pi_{V_1 V_2}$ affect measurable quantities. 
Up to order $p^2$,  the physical combinations are the 3 Peskin--Takeuchi oblique parameters: 
\beq
\label{eq:st} 
\alpha S =  
- 4 {g_L g_Y \over g_L^2 + g_Y^2} \delta \Pi_{3 B}^{(2)},
\quad 
\alpha T  =  {\delta \Pi_{11}^{(0)} - \delta \Pi_{33}^{(0)}  \over m_W^2}, 
\quad 
\alpha U =  {4 g_Y^2 \over g_L^2 + g_Y^2}   \left (\delta \Pi_{11}^{(2)} - \delta \Pi_{33}^{(2)} \right ).  
\eeq 
At order $p^4$ one can define \cite{Barbieri:2004qk} further oblique parameters:  
\beq
\label{eq:wy} 
\alpha V =    m_W^2   \left (\delta \Pi_{11}^{(4)} - \delta \Pi_{33}^{(4)} \right ), 
\quad 
\alpha W  =   - m_W^2   \delta \Pi_{33}^{(4)}, 
\quad 
\alpha X  =   -  m_W^2   \delta \Pi_{3B}^{(4)}, 
\quad 
\alpha Y  =   -  m_W^2  \delta  \Pi_{BB}^{(4)}.  
\eeq
Compared to Ref.~\cite{Barbieri:2004qk}, we rescaled these parameters by $\alpha = e^2/4\pi$.

Now we want to relate the oblique parameters defined above to the vertex corrections and 4-fermion operators in the EFT Lagrangian of \eref{leff}. 
In universal theories, by definition, the new physics effects in the EFT Lagrangian can be represented by only bosonic operators with  $D>4$. 
These operators may lead to corrections to the gauge boson propagators, 
and one can relate their Wilson coefficients to the oblique parameters in \eref{st} and \eref{wy}. 
It turns out that, in an EFT with operators up to $D$=6, only the parameters $S$, $T$, $W$, $Y$ can be generated at tree level~\cite{Barbieri:2004qk}. 
Using field redefinitions and integration by parts we can  get rid of the corrections to the gauge boson propagators, trading them for flavor diagonal vertex corrections and flavor conserving four-fermion operators as in our \eref{leff}.  
Completing this procedure,  we obtain the following map: 
\bea
\label{eq:UNI_STWYtoHB}
\delta g^{Zf_I} &=&   \alpha  \left \{ 
T^3_f { T - W - {g_Y^2\over g_L^2}  Y \over 2}
+ Q_f { 
 2 g_Y^2  T  - (g_L^2 + g_Y^2)  S  + 2 g_Y^2 W + {2 g_Y^2 (2 g_L^2 - g_Y^2) \over g_L^2}  Y   \over 4 (g_L^2 - g_Y^2)} \right \}, 
\nnl 
\,  [c_{\ell \ell}]_{IIJJ} &= &   \alpha  \left [ W -   {g_Y^2\over g_L^2}  Y \right ], 
\quad
 [c_{\ell \ell}]_{IJJI} =  - 2 \alpha W ,  \qquad I < J, 
 \nnl 
\,  [c_{\ell \ell}]_{IIII} &= &  -  \alpha  \left [ W +   {g_Y^2\over g_L^2}  Y \right ], 
\nnl 
 \, [c_{\ell e}]_{IIJJ} &= &   -  {2 g_Y^2 \over g_L^2} \alpha Y , 
     \qquad
  \, [c_{e e}]_{IIJJ} =    -  {4 g_Y^2 \over g_L^2} \alpha Y, 
 \nnl
  \, [c_{\ell q}']_{II JJ}  &=&   -  \alpha  W  ,
  \nnl 
   \qquad
  \, [c_{\ell q}]_{IIJJ}  &=&   {g_Y^2 \over 3 g_L^2} \alpha Y ,
     \qquad 
  \, [c_{\ell u}]_{IIJJ}  =   {4 g_Y^2 \over 3 g_L^2} \alpha Y ,
     \qquad 
  \, [c_{\ell d}]_{IIJJ}  =  - {2 g_Y^2 \over 3 g_L^2} \alpha Y ,
   \nnl 
  \, [c_{e q}]_{IIJJ}  &=&   {2 g_Y^2 \over 3 g_L^2} \alpha Y ,
     \qquad 
  \, [c_{e u}]_{IIJJ}  = {8  g_Y^2 \over 3 g_L^2} \alpha Y ,
     \qquad 
  \, [c_{e d}]_{IIJJ}  =  - {4 g_Y^2 \over 3 g_L^2} \alpha Y .
\eea   
Thus, the oblique parameters  $S$, $T$, $W$, $Y$ correspond to a special pattern of vertex corrections and 4-fermion operators \cite{Wells:2015uba}. 
In our language, the theory can is universal if the pattern of vertex corrections and 4-fermion operators can be matched to  that in \eref{UNI_STWYtoHB}. 
 Note that both leptonic and quark operators are necessarily present in universal theories. 
 Our formalism is more general and applies to a large class of models,  as we don't need to make any assumptions about the pattern of vertex corrections or 4-fermion operators. 
None of the specific models discussed in \sref{models} is universal, and cannot be properly  described by the oblique parameters. 

The current constraints on the oblique parameters are
\beq
\label{eq:styw}
\bvec
S \\ T \\ W \\ Y 
\evec 
= \bvec -0.04 \pm 0.11 \\ 0.05 \pm 0.08 \\ -0.09 \pm 0.08 \\ -0.04 \pm 0.09 \evec, 
\qquad 
\rho = \left(
\begin{array}{cccc}
 1. & 0.88 & 0.57 & 0.08 \\
 . & 1. & 0.3 & 0.04 \\
 . & . & 1. & -0.4 \\
 . & . & . & 1. \\
\end{array}
\right). 
\eeq 
These constraints are dominated by the leptonic observables discussed in this paper.  
To derive \eref{styw},   we also used the experimental input from  the hadronic observables  $e^+ e^- \to q \bar q, b \bar b, c \bar c$ in LEP-1 and LEP-2 \cite{ALEPH:2005ab,Schael:2013ita} and  atomic parity violation \cite{Wood:1997zq}. 
If, instead, we plugged  in \eref{UNI_STWYtoHB} directly in the  likelihood function of \eref{chi2}, 
the result would be very similar as in \eref{styw}, up to a small $\cO(0.05)$ shift of the central values.
Setting $W=Y=0$ one obtains the constraints on $S$ and $T$ alone:
$S = 0.06 \pm 0.08$,  $T = 0.09 \pm 0.07$ with the correlation coefficient $+0.92$, 
which is very close to the result in Ref.~\cite{Baak:2014ora} using the Z-pole and $m_W$ measurements alone.  

\section{Lepton flavor violating operators}
\label{app:lfv}

Among four-lepton operators $[O_{\ell \ell}]_{IJKL}$, $[O_{e e}]_{IJKL}$, $ [O_{\ell e}]_{IJKL}$ with general flavor indices there exist 66 complex ones that violate lepton flavor. 
Moreover,  9 complex lepton flavor violating vertex corrections $[\delta g^{Ze}_{L,R}]_{IJ}$, $[\delta g^{We}_{L}]_{IJ}$ with $I \neq J$ may arise from $D$=6 operators in the EFT.   
These operators do not interfere with the SM and thus, at the leading order,  they do not affect the constraints on flavor conserving operators discussed in \sref{fit}.  
In this appendix, for completeness, we review experimental constraints on some lepton flavor violating vertex corrections and four-lepton operators.
See also \cite{Crivellin:2013hpa, Feruglio:2015yua,Pruna:2015jhf} for recent reviews. 

\subsection{From lepton flavor violating Z decays}

Lepton flavor violating Z boson vertices can be probed by  on-shell Z decays at LEP and the LHC, as recently discussed  Ref.~\cite{Efrati:2015eaa}.
The current experimental limits are:  

 \begin{center}
 \begin{tabular}{|c|c|c|c|c|}
 \hline
{\color{blue}{Observable}} & {\color{blue}{95\% CL limit}}   &   {\color{blue}{Ref.}}  
 \\  \hline  \hline
 ${\rm Br} (Z \to e \mu)$  & $7.5 \times 10^{-7}$ &\cite{Aad:2014bca}  \\ \hline 
 ${\rm Br} (Z \to e \tau)$  & $9.8 \times 10^{-6}$ &\cite{Akers:1995gz}  \\ \hline  
 ${\rm Br} (Z \to \mu \tau)$  & $1.2 \times 10^{-5}$ &\cite{Abreu:1996mj}  \\ \hline 
\end{tabular}
\end{center}

At tree level, this translates to the following constraints on the  vertex corrections:
\bea \label{eq:ZLFV}
\sqrt{|[\delta g_L^{Ze}]_{12}|^2 + | [\delta g_R^{Ze}]_{12}|^2} &<  & 1.2 \times 10^{-3},
\nnl
\sqrt{|[\delta g_L^{Ze}]_{13}|^2 + | [\delta g_R^{Ze}]_{13}|^2} &<  & 4.3 \times 10^{-3},
\nnl
\sqrt{|[\delta g_L^{Ze}]_{23}|^2 + | [\delta g_R^{Ze}]_{23}|^2} &<  & 4.8 \times 10^{-3}. 
\eea
%

\subsection{From lepton flavor violating lepton decay}

Searches for lepton flavor violating muon and tau decays have, so far,  all given negative results and set tight constraints on lepton flavor violating operators. In what follows, we perform a tree-level computation, neglecting the masses of the daughter leptons.

The 90\% CL constraints on the branching ratios given by PDG \cite{Agashe:2014kda} are:
\begin{equation}
\begin{tabular}{|l|c|}
   \hline
 \color{blue}   Decay mode & \color{blue} 90\% CL bound \\
    \hline
  $\mu^- \rightarrow e^- \bar{\nu}_\mu \nu_e $  & 1.2 \%\\
   \hline
 $\mu^- \rightarrow e^- e^+ e^- $   & $1.0 \times 10^{-12}$\\
   \hline
 $\tau^- \rightarrow e^- e^+ e^- $   & $2.7 \times 10^{-8}$\\
    \hline
 $\tau^- \rightarrow \mu^- \mu^+ \mu^- $   & $2.1 \times 10^{-8}$\\
   \hline
 $\tau^- \rightarrow \mu^- e^+ e^- $   & $1.8 \times 10^{-8}$\\
   \hline
 $\tau^- \rightarrow e^- \mu^+ \mu^- $   & $2.7 \times 10^{-8}$\\
   \hline
 $\tau^- \rightarrow e^+ \mu^- \mu^- $    & $1.7 \times 10^{-8}$\\
   \hline
 $\tau^- \rightarrow \mu^+ e^- e^- $    & $1.5 \times 10^{-8}$\\
   \hline
 \end{tabular}
\end{equation}

\bi
\item
$\mu^- \rightarrow e^- \bar{\nu}_\mu \nu_e $\\
This process can be induced by the operators $[O_{\ell\ell}]_{1212}$ and $[O_{\ell e}]_{1212}$. We get, at 90\% CL:
\begin{equation}
\sqrt{4|[c_{\ell\ell}]_{1212}|^2 + |[c_{\ell e}]_{1212}|^2} < 0.219. 
\end{equation}

\item
$\mu^- \rightarrow e^- e^+ e^- $\\
This process can be induced by the operators $[O_{\ell\ell}]_{1112},[O_{\ell e}]_{1112},[O_{\ell e}]_{1211}$ and $[O_{ee}]_{1112}$ but also by vertex corrections.
At 90\% CL:
\begin{eqnarray}
\left\lbrace 2\left| [c_{\ell\ell}]_{1112}+4 g_{L,SM}^{Ze}[\delta g_L^{Ze}]_{12} \right|^2 + 2\left| [c_{ee}]_{1112}+ 4 g_{R,SM}^{Ze}[\delta g_R^{Ze}]_{12} \right|^2 \right. \\ +
\left. \left| [c_{\ell e}]_{1112}+[c_{\ell e}]_{1211}+4 g_{R,SM}^{Ze}[\delta g_L^{Ze}]_{12}+4 g_{L,SM}^{Ze}[\delta g_R^{Ze}]_{12} \right|^2 \right\rbrace ^{1/2} \nonumber\\
< 2.0 \times 10^{-6} \nonumber .
\end{eqnarray}

\item
$\tau^- \rightarrow e^- e^+ e^- $\\
We get, at 90\% CL:
\begin{eqnarray}
\left\lbrace 2\left| [c_{\ell\ell}]_{1113}+4 g_{L,SM}^{Ze}[\delta g_L^{Ze}]_{13} \right|^2 + 2\left| [c_{ee}]_{1113}+ 4 g_{R,SM}^{Ze}[\delta g_R^{Ze}]_{13} \right|^2 \right. \\
+\left. \left| [c_{\ell e}]_{1113}+[c_{\ell e}]_{1311}+4 g_{R,SM}^{Ze}[\delta g_L^{Ze}]_{13}+4 g_{L,SM}^{Ze}[\delta g_R^{Ze}]_{13} \right|^2 \right\rbrace ^{1/2} \nonumber\\
< 7.8 \times 10^{-4} \nonumber . 
\end{eqnarray}

\item
$\tau^- \rightarrow \mu^- \mu^+ \mu^- $\\
We get, at 90\% CL:
\begin{eqnarray}
\left\lbrace 2\left| [c_{\ell\ell}]_{2223}+4 g_{L,SM}^{Ze}[\delta g_L^{Ze}]_{23} \right|^2 + 2\left| [c_{ee}]_{2223}+ 4 g_{R,SM}^{Ze}[\delta g_R^{Ze}]_{23} \right|^2 \right. \nonumber\\+
\left. \left| [c_{\ell e}]_{2223}+[c_{\ell e}]_{2322}+4 g_{R,SM}^{Ze}[\delta g_L^{Ze}]_{23}+4 g_{L,SM}^{Ze}[\delta g_R^{Ze}]_{23} \right|^2 \right\rbrace ^{1/2} \nonumber\\
< 6.9 \times 10^{-4}  . 
\end{eqnarray}

\item
$\tau^- \rightarrow \mu^- e^+ e^- $\\
This process can be induced by the operators $[O_{\ell\ell}]_{1123},[O_{\ell\ell}]_{1321},[O_{\ell e}]_{1123},[O_{\ell e}]_{1321},[O_{\ell e}]_{2113},[O_{\ell e}]_{2311}$ and $[O_{ee}]_{1123}$ but also by vertex corrections
At 90\% CL:
\begin{eqnarray}
\left\lbrace |[c_{\ell\ell}]_{1123}+[c_{\ell\ell}]_{1321}+4 g_{L,SM}^{Ze}[\delta g_L^{Ze}]_{23}|^2+
|[c_{ee}]_{1123}+4 g_{R,SM}^{Ze}[\delta g_R^{Ze}]_{23}|^2\right. + \nonumber\\
\left. |[c_{\ell e}]_{1123}+[c_{\ell e}]_{1321}+4 g_{L,SM}^{Ze}[\delta g_R^{Ze}]_{23}|^2+
|[c_{\ell e}]_{2311}+[c_{\ell e}]_{2113}+4 g_{R,SM}^{Ze}[\delta g_L^{Ze}]_{23}|^2\right\rbrace^{1/2} \nonumber\\
< 6.4 \times 10^{-4}. 
\end{eqnarray}

\item
$\tau^- \rightarrow e^- \mu^+ \mu^- $\\
We get, at 90\% CL:
\begin{eqnarray}
\left\lbrace |[c_{\ell\ell}]_{1322}+[c_{\ell\ell}]_{1223}+4 g_{L,SM}^{Ze}[\delta g_L^{Ze}]_{13}|^2+
|[c_{ee}]_{1223}+4 g_R^{Ze}[\delta g_{R,SM}^{Ze}]_{13}|^2\right. + \nonumber\\
\left. |[c_{\ell e}]_{2213}+[c_{\ell e}]_{2312}+4 g_{L,SM}^{Ze}[\delta g_R^{Ze}]_{13}|^2+
|[c_{\ell e}]_{1322}+[c_{\ell e}]_{1223}+4 g_{R,SM}^{Ze}[\delta g_L^{Ze}]_{13}|^2\right\rbrace^{1/2} \nonumber\\
< 7.8 \times 10^{-4}.
\end{eqnarray}

\item
$\tau^- \rightarrow e^+ \mu^- \mu^- $\\
This process can be induced by the operators $[O_{\ell\ell}]_{2123},[O_{\ell e}]_{2123},[O_{\ell e}]_{2321}$ and $[O_{ee}]_{2123}$ but is not affected at first order by vertex corrections.
At 90\% CL, we have:
\begin{equation}
\sqrt{2|[c_{\ell\ell}]_{2123}|^2+2|[c_{ee}]_{2123}|^2+|[c_{\ell e}]_{2123}+[c_{\ell e}]_{2321}|^2}<6.2\times 10^{-4}.
\end{equation}

\item
$\tau^- \rightarrow \mu^+ e^- e^- $\\
We get, at 90\% CL:
\begin{equation}
\sqrt{2|[c_{\ell\ell}]_{1213}|^2+2|[c_{ee}]_{1213}|^2+|[c_{\ell e}]_{1213}+[c_{\ell e}]_{1312}|^2}<5.8\times 10^{-4}.
\end{equation}
\ei


\subsection{From lepton decay parameters}
In experiments studying lepton decays $\ell \rightarrow \ell_1 \nu \bar \nu$ such as Ref.~\cite{Stahl:1999ui}, the two emitted neutrinos are not detected and are assumed to conserve lepton flavor. A more general analysis of muon decay allowing lepton number violation was presented in Ref.~\cite{Langacker:1988cm}. The authors show that there is a one-to-one correspondence between the form factors $g^\gamma_{\epsilon\mu}$ defined e.g. in Ref.~\cite{Stahl:1999ui} in the lepton flavor conserving case and combinations of parameters in the lepton number violating case.

In our D=6 EFT framework, for the decay $\ell_I \rightarrow \ell_J \nu \bar\nu$, this correspondence is:
\begin{eqnarray}
|g^S_{RR}|^2\rightarrow \sum_{k \ge l} |[c_{\ell e}]_{klJI}|^2, \nnl
|g^V_{RR}|^2\rightarrow \sum_{k < l} |[c_{\ell e}]_{klJI}|^2, \nnl 
|g^S_{LL}|^2\rightarrow \sum_{k > l} |a_{lk}[c_{\ell \ell}]_{JIlk}|^2,
\end{eqnarray}
where $a_{JI}=2$ and $a_{kl}=1$ in all other cases.

The limits given by PDG \cite{Agashe:2014kda} are:
\begin{eqnarray}
|g^S_{RR}|<0.035, \qquad & |g^V_{RR}|< 0.017, \qquad  |g^S_{LL}|< 0.550, & \mbox{for} \; \mu^-\rightarrow e^- \bar \nu \nu \; \mbox{at 90\% CL} . 
\nnl 
|g^S_{RR}|<0.70, \qquad & |g^V_{RR}|< 0.17, \qquad  |g^S_{LL}|< 2.01, & \mbox{for} \; \tau^-\rightarrow e^- \bar \nu \nu \; \mbox{at 95\% CL}, 
\nnl
|g^S_{RR}|<0.72,  \qquad & |g^V_{RR}|< 0.18, \qquad  |g^S_{LL}|< 2.01, & \mbox{for} \; \tau^-\rightarrow \mu^- \bar \nu \nu \; \mbox{at 95\% CL}. 
\end{eqnarray}
The constraints are quite weak for tau decays parameters, but give constraints better than $0.1$ for Wilson coefficient of nine four-lepton operators. Explicitly they are:

\begin{eqnarray}
\sqrt{|[c_{\ell e}]_{1112}|^2+|[c_{\ell e}]_{2112}|^2+|[c_{\ell e}]_{3112}|^2+|[c_{\ell e}]_{2212}|^2+|[c_{\ell e}]_{3212}|^2+|[c_{\ell e}]_{3312}|^2}<0.035,  \nnl
\sqrt{|[c_{\ell e}]_{1212}|^2+|[c_{\ell e}]_{1312}|^2+|[c_{\ell e}]_{2312}|^2}<0.017,
\end{eqnarray}
at 90\% CL.

\newpage
\pagestyle{empty} 
\begin{landscape}
\section{Correlation Matrix}
\label{app:correlation}
{\scriptsize
\bea 
\label{eq:fullmonty_rho}
&& \hspace{-1.5cm} \quad \rho =  \nnl 
&& \hspace{-1.5cm} 
\left(
\begin{array}{cccccccccccccccccccccc}
 1. & -0.12 & -0.63 & -0.1 & -0.03 & 0.01 & 0.07 & -0.06 & -0.04 & 0. & -0.01 & 0. & 0.7 & 0.03 & 0.03 & -0.03 & -0.47 & 0.2 & 0. & 0. & 0. & -0.51 \\
 . & 1. & -0.56 & -0.11 & -0.04 & 0.01 & 0.08 & -0.06 & -0.04 & 0. & -0.01 & 0. & 0.63 & -0.28 & -0.24 & 0.24 & -0.15 & -0.77 & 0.01 & 0. & 0. & 0.13 \\
 . & . & 1. & -0.1 & -0.03 & 0.01 & 0.07 & -0.05 & -0.04 & 0. & -0.01 & 0. & -0.9 & 0.16 & 0.13 & -0.14 & 0.43 & 0.58 & 0. & 0. & 0. & 0.41 \\
 . & . & . & 1. & -0.1 & -0.07 & 0.17 & -0.05 & 0.03 & -0.09 & 0.04 & 0.08 & -0.16 & 0.06 & 0.03 & -0.03 & 0.05 & -0.21 & 0. & -0.01 & 0. & -0.13 \\
 . & . & . & . & 1. & 0.07 & -0.06 & 0.9 & -0.04 & 0.01 & 0. & -0.01 & -0.05 & -0.04 & -0.03 & 0.04 & 0.08 & -0.07 & 0. & 0. & 0. & -0.04 \\
 . & . & . & . & . & 1. & 0.02 & -0.03 & 0.41 & 0.01 & 0. & 0. & 0.01 & -0.01 & -0.01 & 0.01 & 0. & 0.02 & 0.01 & -0.01 & -0.01 & 0.01 \\
 . & . & . & . & . & . & 1. & -0.08 & -0.04 & -0.08 & -0.05 & 0.08 & 0.12 & -0.02 & -0.04 & 0.04 & -0.06 & 0.15 & 0. & 0.01 & 0. & 0.09 \\
 . & . & . & . & . & . & . & 1. & 0.04 & 0.01 & 0. & -0.01 & -0.09 & -0.03 & -0.02 & 0.02 & 0.09 & -0.12 & 0. & 0. & 0. & -0.07 \\
 . & . & . & . & . & . & . & . & 1. & 0. & 0. & 0. & -0.06 & 0.02 & 0.01 & -0.01 & 0.03 & -0.08 & 0. & 0.02 & 0. & -0.05 \\
 . & . & . & . & . & . & . & . & . & 1. & -0.53 & -0.08 & 0.01 & 0. & 0. & 0. & 0. & 0.01 & 0. & 0. & 0. & 0.01 \\
 . & . & . & . & . & . & . & . & . & . & 1. & -0.52 & -0.01 & 0. & 0. & 0. & 0.01 & -0.02 & 0. & 0. & 0. & -0.01 \\
 . & . & . & . & . & . & . & . & . & . & . & 1. & 0. & 0. & 0. & 0. & 0. & 0. & 0. & 0. & 0. & 0. \\
 . & . & . & . & . & . & . & . & . & . & . & . & 1. & -0.18 & -0.15 & 0.15 & -0.48 & -0.39 & 0. & 0. & 0. & -0.3 \\
 . & . & . & . & . & . & . & . & . & . & . & . & . & 1. & 0.04 & -0.04 & -0.77 & 0.21 & 0. & 0. & 0. & -0.04 \\
 . & . & . & . & . & . & . & . & . & . & . & . & . & . & 1. & -0.98 & 0.05 & 0.18 & 0. & 0. & 0. & -0.03 \\
 . & . & . & . & . & . & . & . & . & . & . & . & . & . & . & 1. & -0.06 & -0.19 & 0. & 0. & 0. & 0.03 \\
 . & . & . & . & . & . & . & . & . & . & . & . & . & . & . & . & 1. & 0.06 & 0. & 0. & 0. & 0.22 \\
 . & . & . & . & . & . & . & . & . & . & . & . & . & . & . & . & . & 1. & -0.01 & 0. & 0. & 0.01 \\
 . & . & . & . & . & . & . & . & . & . & . & . & . & . & . & . & . & . & 1. & 0.01 & -1. & 0. \\
 . & . & . & . & . & . & . & . & . & . & . & . & . & . & . & . & . & . & . & 1. & -0.01 & 0. \\
 . & . & . & . & . & . & . & . & . & . & . & . & . & . & . & . & . & . & . & . & 1. & 0. \\
 . & . & . & . & . & . & . & . & . & . & . & . & . & . & . & . & . & . & . & . & . & 1. \\
\end{array}
\right)
\nnl 
\eea }
\end{landscape}

\bibliographystyle{JHEP}
\bibliography{fourpaper}

\providecommand{\href}[2]{#2}\begingroup\raggedright\begin{thebibliography}{10}

\bibitem{Buchmuller:1985jz}
W.~Buchmuller and D.~Wyler, {\it {Effective Lagrangian Analysis of New
  Interactions and Flavor Conservation}},  {\em Nucl.Phys.} {\bf B268} (1986)
  621--653.

\bibitem{Han:2004az}
Z.~Han and W.~Skiba, {\it {Effective theory analysis of precision electroweak
  data}},  {\em Phys.Rev.} {\bf D71} (2005) 075009,
  [\href{http://arxiv.org/abs/hep-ph/0412166}{{\tt hep-ph/0412166}}].

\bibitem{Han:2005pr}
Z.~Han, {\it {Electroweak constraints on effective theories with U(2) x (1)
  flavor symmetry}},  {\em Phys. Rev.} {\bf D73} (2006) 015005,
  [\href{http://arxiv.org/abs/hep-ph/0510125}{{\tt hep-ph/0510125}}].

\bibitem{Barbieri:2004qk}
R.~Barbieri, A.~Pomarol, R.~Rattazzi, and A.~Strumia, {\it {Electroweak
  symmetry breaking after LEP-1 and LEP-2}},  {\em Nucl.Phys.} {\bf B703}
  (2004) 127--146, [\href{http://arxiv.org/abs/hep-ph/0405040}{{\tt
  hep-ph/0405040}}].

\bibitem{Grojean:2006nn}
C.~Grojean, W.~Skiba, and J.~Terning, {\it {Disguising the oblique
  parameters}},  {\em Phys.Rev.} {\bf D73} (2006) 075008,
  [\href{http://arxiv.org/abs/hep-ph/0602154}{{\tt hep-ph/0602154}}].

\bibitem{Cacciapaglia:2006pk}
G.~Cacciapaglia, C.~Csaki, G.~Marandella, and A.~Strumia, {\it {The Minimal Set
  of Electroweak Precision Parameters}},  {\em Phys.Rev.} {\bf D74} (2006)
  033011, [\href{http://arxiv.org/abs/hep-ph/0604111}{{\tt hep-ph/0604111}}].

\bibitem{Pomarol:2013zra}
A.~Pomarol and F.~Riva, {\it {Towards the Ultimate SM Fit to Close in on Higgs
  Physics}},  {\em JHEP} {\bf 1401} (2014) 151,
  [\href{http://arxiv.org/abs/1308.2803}{{\tt arXiv:1308.2803}}].

\bibitem{Elias-Miro:2013mua}
J.~Elias-Miro, J.~Espinosa, E.~Masso, and A.~Pomarol, {\it {Higgs windows to
  new physics through d=6 operators: constraints and one-loop anomalous
  dimensions}},  {\em JHEP} {\bf 1311} (2013) 066,
  [\href{http://arxiv.org/abs/1308.1879}{{\tt arXiv:1308.1879}}].

\bibitem{Dumont:2013wma}
B.~Dumont, S.~Fichet, and G.~von Gersdorff, {\it {A Bayesian view of the Higgs
  sector with higher dimensional operators}},  {\em JHEP} {\bf 1307} (2013)
  065, [\href{http://arxiv.org/abs/1304.3369}{{\tt arXiv:1304.3369}}].

\bibitem{Chen:2013kfa}
C.-Y. Chen, S.~Dawson, and C.~Zhang, {\it {Electroweak Effective Operators and
  Higgs Physics}},  {\em Phys. Rev.} {\bf D89} (2014), no.~1 015016,
  [\href{http://arxiv.org/abs/1311.3107}{{\tt arXiv:1311.3107}}].

\bibitem{deBlas:2013qqa}
J.~de~Blas, M.~Chala, and J.~Santiago, {\it {Global Constraints on Lepton-Quark
  Contact Interactions}},  {\em Phys. Rev.} {\bf D88} (2013) 095011,
  [\href{http://arxiv.org/abs/1307.5068}{{\tt arXiv:1307.5068}}].

\bibitem{Willenbrock:2014bja}
S.~Willenbrock and C.~Zhang, {\it {Effective Field Theory Beyond the Standard
  Model}},  \href{http://arxiv.org/abs/1401.0470}{{\tt arXiv:1401.0470}}.

\bibitem{Gupta:2014rxa}
R.~S. Gupta, A.~Pomarol, and F.~Riva, {\it {BSM Primary Effects}},  {\em
  Phys.Rev.} {\bf D91} (2015), no.~3 035001,
  [\href{http://arxiv.org/abs/1405.0181}{{\tt arXiv:1405.0181}}].

\bibitem{Masso:2014xra}
E.~Masso, {\it {An Effective Guide to Beyond the Standard Model Physics}},
  {\em JHEP} {\bf 1410} (2014) 128, [\href{http://arxiv.org/abs/1406.6376}{{\tt
  arXiv:1406.6376}}].

\bibitem{deBlas:2014ula}
J.~de~Blas, M.~Ciuchini, E.~Franco, D.~Ghosh, S.~Mishima, et~al., {\it {Global
  Bayesian Analysis of the Higgs-boson Couplings}},
  \href{http://arxiv.org/abs/1410.4204}{{\tt arXiv:1410.4204}}.

\bibitem{Ciuchini:2014dea}
M.~Ciuchini, E.~Franco, S.~Mishima, M.~Pierini, L.~Reina, et~al., {\it {Update
  of the electroweak precision fit, interplay with Higgs-boson signal strengths
  and model-independent constraints on new physics}},
  \href{http://arxiv.org/abs/1410.6940}{{\tt arXiv:1410.6940}}.

\bibitem{Ellis:2014huj}
J.~Ellis, V.~Sanz, and T.~You, {\it {The Effective Standard Model after LHC Run
  I}},  \href{http://arxiv.org/abs/1410.7703}{{\tt arXiv:1410.7703}}.

\bibitem{Falkowski:2014tna}
A.~Falkowski and F.~Riva, {\it {Model-independent precision constraints on
  dimension-6 operators}},  {\em JHEP} {\bf 1502} (2015) 039,
  [\href{http://arxiv.org/abs/1411.0669}{{\tt arXiv:1411.0669}}].

\bibitem{delAguila:2014soa}
F.~del Aguila, M.~Chala, J.~Santiago, and Y.~Yamamoto, {\it {Collider limits on
  leptophilic interactions}},  {\em JHEP} {\bf 03} (2015) 059,
  [\href{http://arxiv.org/abs/1411.7394}{{\tt arXiv:1411.7394}}].

\bibitem{Corbett:2015ksa}
T.~Corbett, O.~J.~P. Eboli, D.~Goncalves, J.~Gonzalez-Fraile, T.~Plehn, and
  M.~Rauch, {\it {The Higgs Legacy of the LHC Run I}},  {\em JHEP} {\bf 08}
  (2015) 156, [\href{http://arxiv.org/abs/1505.05516}{{\tt arXiv:1505.05516}}].

\bibitem{Efrati:2015eaa}
A.~Efrati, A.~Falkowski, and Y.~Soreq, {\it {Electroweak constraints on
  flavorful effective theories}},  {\em JHEP} {\bf 07} (2015) 018,
  [\href{http://arxiv.org/abs/1503.07872}{{\tt arXiv:1503.07872}}].

\bibitem{Gonzalez-Alonso:2015bha}
M.~Gonzalez-Alonso, A.~Greljo, G.~Isidori, and D.~Marzocca, {\it {Electroweak
  bounds on Higgs pseudo-observables and $h \to 4 \ell$ decays}},  {\em Eur.
  Phys. J.} {\bf C75} (2015) 341, [\href{http://arxiv.org/abs/1504.04018}{{\tt
  arXiv:1504.04018}}].

\bibitem{Buckley:2015nca}
A.~Buckley, C.~Englert, J.~Ferrando, D.~J. Miller, L.~Moore, M.~Russell, and
  C.~D. White, {\it {A global fit of top quark effective theory to data}},
  \href{http://arxiv.org/abs/1506.08845}{{\tt arXiv:1506.08845}}.

\bibitem{deBlas:2015aea}
J.~de~Blas, M.~Chala, and J.~Santiago, {\it {Renormalization Group Constraints
  on New Top Interactions from Electroweak Precision Data}},  {\em JHEP} {\bf
  09} (2015) 189, [\href{http://arxiv.org/abs/1507.00757}{{\tt
  arXiv:1507.00757}}].

\bibitem{Falkowski:2015fla}
A.~Falkowski, {\it {Effective field theory approach to LHC Higgs data}},
  \href{http://arxiv.org/abs/1505.00046}{{\tt arXiv:1505.00046}}.

\bibitem{delAguila:2015vza}
F.~del Aguila, M.~Chala, J.~Santiago, and Y.~Yamamoto, {\it {Four and
  two-lepton signals of leptophilic gauge interactions at large colliders}},
  {\em PoS} {\bf CORFU2014} (2015) 109,
  [\href{http://arxiv.org/abs/1505.00799}{{\tt arXiv:1505.00799}}].

\bibitem{Wells:2015eba}
J.~D. Wells and Z.~Zhang, {\it {Status and prospects of precision analyses with
  $e^+e^-\to W^+W^-$}},  \href{http://arxiv.org/abs/1507.01594}{{\tt
  arXiv:1507.01594}}.

\bibitem{Berthier:2015gja}
L.~Berthier and M.~Trott, {\it {Consistent constraints on the Standard Model
  Effective Field Theory}},  \href{http://arxiv.org/abs/1508.05060}{{\tt
  arXiv:1508.05060}}.

\bibitem{Ellis:2015sca}
J.~Ellis and T.~You, {\it {Sensitivities of Prospective Future e+e- Colliders
  to Decoupled New Physics}},  \href{http://arxiv.org/abs/1510.04561}{{\tt
  arXiv:1510.04561}}.

\bibitem{Englert:2015hrx}
C.~Englert, R.~Kogler, H.~Schulz, and M.~Spannowsky, {\it {Higgs coupling
  measurements at the LHC}},  \href{http://arxiv.org/abs/1511.05170}{{\tt
  arXiv:1511.05170}}.

\bibitem{Grzadkowski:2010es}
B.~Grzadkowski, M.~Iskrzynski, M.~Misiak, and J.~Rosiek, {\it {Dimension-Six
  Terms in the Standard Model Lagrangian}},  {\em JHEP} {\bf 1010} (2010) 085,
  [\href{http://arxiv.org/abs/1008.4884}{{\tt arXiv:1008.4884}}].

\bibitem{Alonso:2013hga}
R.~Alonso, E.~E. Jenkins, A.~V. Manohar, and M.~Trott, {\it {Renormalization
  Group Evolution of the Standard Model Dimension Six Operators III: Gauge
  Coupling Dependence and Phenomenology}},  {\em JHEP} {\bf 1404} (2014) 159,
  [\href{http://arxiv.org/abs/1312.2014}{{\tt arXiv:1312.2014}}].

\bibitem{Dreiner:2008tw}
H.~K. Dreiner, H.~E. Haber, and S.~P. Martin, {\it {Two-component spinor
  techniques and Feynman rules for quantum field theory and supersymmetry}},
  {\em Phys.Rept.} {\bf 494} (2010) 1--196,
  [\href{http://arxiv.org/abs/0812.1594}{{\tt arXiv:0812.1594}}].

\bibitem{HB}
{\bf LHC Higgs Cross Section Working Group 2} Collaboration, {\it { Higgs
  Basis: Proposal for an EFT basis choice for LHC HXSWG}},
  \href{http://arxiv.org/abs/LHCHXSWG-INT-2015-001}{{\tt
  LHCHXSWG-INT-2015-001}}.

\bibitem{ALEPH:2005ab}
{\bf ALEPH, DELPHI, L3, OPAL, SLD, LEP Electroweak Working Group, SLD
  Electroweak Group, SLD Heavy Flavour Group} Collaboration, S.~Schael et~al.,
  {\it {Precision electroweak measurements on the $Z$ resonance}},  {\em
  Phys.Rept.} {\bf 427} (2006) 257--454,
  [\href{http://arxiv.org/abs/hep-ex/0509008}{{\tt hep-ex/0509008}}].

\bibitem{Schael:2013ita}
{\bf ALEPH, DELPHI, L3, OPAL, LEP Electroweak Working Group} Collaboration,
  S.~Schael et~al., {\it {Electroweak Measurements in Electron-Positron
  Collisions at W-Boson-Pair Energies at LEP}},  {\em Phys.Rept.} {\bf 532}
  (2013) 119--244, [\href{http://arxiv.org/abs/1302.3415}{{\tt
  arXiv:1302.3415}}].

\bibitem{Group:2012gb}
{\bf CDF and D0} Collaboration, T.~E.~W. Group, {\it {2012 Update of the
  Combination of CDF and D0 Results for the Mass of the W Boson}},
  \href{http://arxiv.org/abs/1204.0042}{{\tt arXiv:1204.0042}}.

\bibitem{Dorenbosch:1988is}
{\bf CHARM} Collaboration, J.~Dorenbosch et~al., {\it {EXPERIMENTAL RESULTS ON
  NEUTRINO - ELECTRON SCATTERING}},  {\em Z. Phys.} {\bf C41} (1989) 567.
  [Erratum: Z. Phys.C51,142(1991)].

\bibitem{Vilain:1994qy}
{\bf CHARM-II} Collaboration, P.~Vilain et~al., {\it {Precision measurement of
  electroweak parameters from the scattering of muon-neutrinos on electrons}},
  {\em Phys. Lett.} {\bf B335} (1994) 246--252.

\bibitem{Ahrens:1990fp}
L.~A. Ahrens et~al., {\it {Determination of electroweak parameters from the
  elastic scattering of muon-neutrinos and anti-neutrinos on electrons}},  {\em
  Phys. Rev.} {\bf D41} (1990) 3297--3316.

\bibitem{Erler:2013xha}
J.~Erler and S.~Su, {\it {The Weak Neutral Current}},  {\em Prog. Part. Nucl.
  Phys.} {\bf 71} (2013) 119--149, [\href{http://arxiv.org/abs/1303.5522}{{\tt
  arXiv:1303.5522}}].

\bibitem{Auerbach:2001wg}
{\bf LSND} Collaboration, L.~B. Auerbach et~al., {\it {Measurement of electron
  - neutrino - electron elastic scattering}},  {\em Phys. Rev.} {\bf D63}
  (2001) 112001, [\href{http://arxiv.org/abs/hep-ex/0101039}{{\tt
  hep-ex/0101039}}].

\bibitem{Abe:2012gx}
{\bf T2K} Collaboration, K.~Abe et~al., {\it {First Muon-Neutrino Disappearance
  Study with an Off-Axis Beam}},  {\em Phys. Rev.} {\bf D85} (2012) 031103,
  [\href{http://arxiv.org/abs/1201.1386}{{\tt arXiv:1201.1386}}].

\bibitem{Anthony:2005pm}
{\bf SLAC E158} Collaboration, P.~L. Anthony et~al., {\it {Precision
  measurement of the weak mixing angle in Moller scattering}},  {\em Phys. Rev.
  Lett.} {\bf 95} (2005) 081601,
  [\href{http://arxiv.org/abs/hep-ex/0504049}{{\tt hep-ex/0504049}}].

\bibitem{Czarnecki:1995fw}
A.~Czarnecki and W.~J. Marciano, {\it {Electroweak radiative corrections to
  polarized Moller scattering asymmetries}},  {\em Phys. Rev.} {\bf D53} (1996)
  1066--1072, [\href{http://arxiv.org/abs/hep-ph/9507420}{{\tt
  hep-ph/9507420}}].

\bibitem{Agashe:2014kda}
{\bf Particle Data Group} Collaboration, K.~A. Olive et~al., {\it {Review of
  Particle Physics}},  {\em Chin. Phys.} {\bf C38} (2014) 090001.

\bibitem{Bouchiat:1957zz}
C.~Bouchiat and L.~Michel, {\it {Theory of $\mu$-Meson Decay with the
  Hypothesis of Nonconservation of Parity}},  {\em Phys. Rev.} {\bf 106} (1957)
  170--172. [,89(1957)].

\bibitem{Gonzalez-Alonso:2014bga}
M.~Gonzalez-Alonso, {\it {TAU2014 Opening Talk}},  {\em Nucl. Part. Phys.
  Proc.} {\bf 260} (2015) 3--11, [\href{http://arxiv.org/abs/1411.4529}{{\tt
  arXiv:1411.4529}}].

\bibitem{martinthesis}
M.~Gonzalez-Alonso, {\it Test del modelo estandar a energias bajas},  {\em Phd
  Thesis} (2010).

\bibitem{Danneberg:2005xv}
N.~Danneberg et~al., {\it {Muon decay: Measurement of the transverse
  polarization of the decay positrons and its implications for the Fermi
  coupling constant and time reversal invariance}},  {\em Phys. Rev. Lett.}
  {\bf 94} (2005) 021802.

\bibitem{Baer:2013cma}
H.~Baer, T.~Barklow, K.~Fujii, Y.~Gao, A.~Hoang, S.~Kanemura, J.~List, H.~E.
  Logan, A.~Nomerotski, M.~Perelstein, et~al., {\it {The International Linear
  Collider Technical Design Report - Volume 2: Physics}},
  \href{http://arxiv.org/abs/1306.6352}{{\tt arXiv:1306.6352}}.

\bibitem{Koratzinos:2013ncw}
M.~Koratzinos et~al., {\it {TLEP: A High-Performance Circular $e^+e^-$ Collider
  to Study the Higgs Boson}},  in {\em {Proceedings, 4th International Particle
  Accelerator Conference (IPAC 2013)}}, p.~TUPME040, 2013.
\newblock \href{http://arxiv.org/abs/1305.6498}{{\tt arXiv:1305.6498}}.

\bibitem{Crivellin:2013hpa}
A.~Crivellin, S.~Najjari, and J.~Rosiek, {\it {Lepton Flavor Violation in the
  Standard Model with general Dimension-Six Operators}},  {\em JHEP} {\bf 04}
  (2014) 167, [\href{http://arxiv.org/abs/1312.0634}{{\tt arXiv:1312.0634}}].

\bibitem{Peskin:1991sw}
M.~E. Peskin and T.~Takeuchi, {\it {Estimation of oblique electroweak
  corrections}},  {\em Phys.Rev.} {\bf D46} (1992) 381--409.

\bibitem{Wells:2015uba}
J.~D. Wells and Z.~Zhang, {\it {Effective theories of universal theories}},
  \href{http://arxiv.org/abs/1510.08462}{{\tt arXiv:1510.08462}}.

\bibitem{Wood:1997zq}
C.~S. Wood, S.~C. Bennett, D.~Cho, B.~P. Masterson, J.~L. Roberts, C.~E.
  Tanner, and C.~E. Wieman, {\it {Measurement of parity nonconservation and an
  anapole moment in cesium}},  {\em Science} {\bf 275} (1997) 1759--1763.

\bibitem{Baak:2014ora}
{\bf Gfitter Group} Collaboration, M.~Baak et~al., {\it {The global electroweak
  fit at NNLO and prospects for the LHC and ILC}},  {\em Eur.Phys.J.} {\bf C74}
  (2014) 3046, [\href{http://arxiv.org/abs/1407.3792}{{\tt arXiv:1407.3792}}].

\bibitem{Feruglio:2015yua}
F.~Feruglio, {\it {Theoretical Aspects of Flavour and CP Violation in the
  Lepton Sector}},  in {\em {27th Rencontres de Blois on Particle Physics and
  Cosmology Blois, France, May 31-June 5, 2015}}, 2015.
\newblock \href{http://arxiv.org/abs/1509.08428}{{\tt arXiv:1509.08428}}.

\bibitem{Pruna:2015jhf}
G.~M. Pruna and A.~Signer, {\it {Lepton-flavour violating decays in theories
  with dimension 6 operators}},  in {\em {Proceedings, GPU Computing in
  High-Energy Physics (GPUHEP2014)}}, 2015.
\newblock \href{http://arxiv.org/abs/1511.04421}{{\tt arXiv:1511.04421}}.

\bibitem{Aad:2014bca}
{\bf ATLAS} Collaboration, G.~Aad et~al., {\it {Search for the lepton flavor
  violating decay $Z \to e \mu$ in pp collisions at $\sqrt{s}=8$~TeV with the
  ATLAS detector}},  {\em Phys.Rev.} {\bf D90} (2014), no.~7 072010,
  [\href{http://arxiv.org/abs/1408.5774}{{\tt arXiv:1408.5774}}].

\bibitem{Akers:1995gz}
{\bf OPAL} Collaboration, R.~Akers et~al., {\it {A Search for lepton flavor
  violating Z0 decays}},  {\em Z.Phys.} {\bf C67} (1995) 555--564.

\bibitem{Abreu:1996mj}
{\bf DELPHI} Collaboration, P.~Abreu et~al., {\it {Search for lepton flavor
  number violating Z0 decays}},  {\em Z.Phys.} {\bf C73} (1997) 243--251.

\bibitem{Stahl:1999ui}
A.~Stahl, {\it {Michel parameters: Averages and interpretation}},  {\em Nucl.
  Phys. Proc. Suppl.} {\bf 76} (1999) 173--181.

\bibitem{Langacker:1988cm}
P.~Langacker and D.~London, {\it {Analysis of Muon Decay With Lepton Number
  Nonconserving Interactions}},  {\em Phys. Rev.} {\bf D39} (1989) 266.

\end{thebibliography}\endgroup

\end{document}